\documentclass[12pt, a4paper]{article}

\usepackage[utf8]{inputenc}
\usepackage[T1]{fontenc}

\usepackage{longtable}

\usepackage{amsmath}
\usepackage{amssymb}
\usepackage{amsfonts}
\usepackage{amsthm} % Load before defining theorem environments
\usepackage{bbm}
\usepackage{mathtools}

\usepackage{graphicx}
\usepackage{booktabs}
\usepackage{float}
\usepackage{caption}
\usepackage{diagbox}
\usepackage{longtable}
\usepackage{multirow}
\usepackage{booktabs}
\usepackage{tikz}
\usetikzlibrary{arrows.meta, positioning}

\usepackage{xcolor}
\usepackage{multicol}
\usepackage{enumitem}
\usepackage{bm}
\usepackage{algorithm}
\usepackage{algpseudocode}

\usepackage{tikz}

\usepackage[most]{tcolorbox}

\usepackage{comment}

\usepackage{url} 

\newcommand{\ssitem}[1][black]{\stepcounter{enumii}\item[\color{1}$\bm{*}$\,\textbf{(\alph{enumii})}]}
\newcommand{\sitem}[1][black]{\stepcounter{enumi}\item[\color{1}$\bm{*}$\,\theenumi.]}
\newcommand{\dsitem}[1][black]{\stepcounter{enumi}\item[\color{1}$\bm{}$\,\theenumi.]}

\setlist[enumerate,1]{label=\textbf{\arabic*.}}
\setlist[enumerate,2]{label=\textbf{\alph*)}}

\newtheoremstyle{noparens}%
    {}{}                  % Space above and below
    {\itshape}{}          % Body font and indentation
    {\bfseries}{}         % Header font and punctuation
    { }                   % Space after theorem header
    {\thmnote{}}        % Optional argument without parentheses

\theoremstyle{noparens}

\theoremstyle{noparens}

\title{Identifying and Quantifying Financial Bubbles \\
with the Hyped Log-Periodic Power Law Model
}

\makeatletter
\def\@fnsymbol#1{\ifcase#1\or *\or **\or  †\or ‡\or §\or ¶\or ††\or ‡‡\fi}
\makeatother

\author{%
  Zheng Cao\thanks{All authors contributed equally},
  Xingran Shao,
  Yuheng Yan,
  Helyette Geman\thanks{Corresponding author} \\
  \texttt{(zcao26, xshao12, yyan75)@jh.edu,hgeman1@jhu.edu} \\
  Department of Applied Mathematics and Statistics\\
  Johns Hopkins University \\
}

\date{}

\begin{document}

\maketitle

\begin{abstract}

%研究超买超卖现象

We propose a novel model, the Hyped Log-Periodic Power Law Model (HLPPL), to the problem of quantifying and detecting financial bubbles, an ever-fascinating one for academics and practitioners alike. Bubble labels are generated using a Log-Periodic Power Law (LPPL) model, sentiment scores, and a hype index we introduced in previous research on NLP forecasting of stock return volatility. Using these tools, a dual-stream transformer model is trained with market data and machine learning methods, resulting in a time series of confidence scores as a Bubble Score. A distinctive feature of our framework is that it captures phases of extreme overpricing and underpricing within a unified structure. 

  % We propose a novel approach to the problem of quantifying and detecting financial bubbles and negative bubbles, an ever fascinating one for academics and practitioners alike. Bubble labels are generated using a Log-Periodic Power Law (LPPL) model, sentiment scores, and a hype index we introduced in previous research on NLP forecasting of shares return volatility. Using these tools, a dual-stream transformer model is trained with market data and machine learning methods, resulting in a time series of confidence scores as a Bubble Score.  A distinctive feature of our framework is that  it captures phases of extreme overpricing and underpricing within a unified structure.

%  A distinctive feature of our framework is that it relies on a single LPPL bubble trajectory to study both super-buy and super-sell phenomena, thereby capturing phases of extreme overpricing and underpricing within a unified structure.

We achieve an average yield of 34.13 percentage annualized return when backtesting U.S. equities during the period 2018 to 2024, while the approach exhibits a remarkable generalization ability across industry sectors. Its conservative bias in predicting bubble periods minimizes false positives, a feature which is especially beneficial for market signaling and decision-making. Overall, this approach utilizes both theoretical and empirical advances for real-time positive and negative bubble identification and measurement with HLPPL signals. 

  % We achieve an average yield of    {34.13} percent annualized return when backtesting U.S. equities during the period 2018 to 2024, while the approach exhibits a remarkable generalization ability across industry sectors. Its conservative bias in predicting bubble periods minimizes false positives, a feature which is especially beneficial for market signaling and decision-making. Overall, this approach utilizes both theoretical and empirical advances for real-time bubble measurement and identification with NLP and LPPL, with practical applications in risk management and portfolio optimization.

\end{abstract}


\newpage

\section{Introduction} \label{section: introduction}

% 背景、研究动机

\hspace{1.5em}The notion of a financial ``bubble" dates back to the early 18th century South Sea Bubble, where inflated company stocks were described as fragile and insubstantial, like air-filled bubbles destined to burst. Sir Isaac Newton, one of history's greatest minds, reportedly lost nearly $£20,000$ in the collapse, lamenting that he could ``calculate the motions of heavenly bodies, but not the madness of people." Earlier episodes such as the Dutch tulip mania were termed ``manias," but the bubble metaphor has since become the dominant way to capture both the rapid inflation and sudden collapse of asset prices.

In financial economics, a bubble refers to a situation where an asset’s price exceeds its fundamental value by a significant margin, driven by expectations of further price increases rather than justified by intrinsic returns.

This research makes several significant contributions to the financial bubble detection and quantification literature. First, we introduce the systematic framework for identifying and quantifying negative bubbles, expanding the traditional focus beyond overvaluation episodes. Second, we demonstrate how behavioral finance indicators can be effectively integrated with the technical Log-Periodic Power Law Model (LPPL) analysis to improve bubble detection accuracy. Third, we develop a practical trading strategy that translates bubble signals into actionable investment decisions, providing empirical evidence of the framework's profitability.

 The remainder of this paper is organized as follows. Section \ref{section: literature review} reviews the relevant literature on financial bubbles, LPPL modeling, Natural Language Processing (NLP) for Finance, and Hype Index. Section \ref{section:BubbleScore}  presents the Hyped Log-Periodic Power Law Model (HLPPL), including the mathematical framework for bidirectional bubble detection and behavioral indicator integration.  Section \ref{section: transformer model} introduces the transformer model and describes the data sources and empirical implementation. Section \ref{section: Trading Strategy and Backtest based on Bubblescore} presents comprehensive backtesting results across multiple assets and time periods. Section \ref{section: Machine Learning Enhanced Trading Strategy} presents the improved model results with Machine Learning (ML) models.
 Section \ref{section: conclusion} concludes with suggestions for future research directions.

 The empirical analysis demonstrates that the HLPPL framework generates superior returns compared to traditional buy-and-hold strategies and pure LPPL approaches. The integration of behavioral indicators significantly improves bubble detection accuracy, while the bidirectional framework captures valuable trading opportunities that would be missed by conventional methods. These findings have important implications for portfolio management, risk assessment, and market timing strategies in both institutional and retail investment contexts.

\section{Literature Review} \label{section: literature review}

\hspace{1.5em}The detection and prediction of financial bubbles have long been central challenges in financial economics, with significant implications for market stability, risk management, and investment decision-making. Traditional approaches to bubble detection often rely on fundamental valuation metrics or statistical anomalies in price movements. Campbell and Shiller (1987) \cite{campbell1987stock} show that stock prices deviating from earnings and dividends may signal bubbles, as traditional detection methods rely mainly on fundamentals or price anomalies. However, these methods frequently fail to capture the complex dynamics underlying speculative episodes. This paper introduces an enhanced framework that integrates the Log-Periodic Power Law Singularity (LPPL) model with machine learning to provide a more comprehensive approach to bubble detection and trading strategy development.
% behavior finance indicators

The Log-Periodic Power Law Singularity (LPPL) model was first introduced by
Johansen, Ledoit, and Sornette (2000)\cite{johansen1999crashes}.
The model captures the dynamics of financial bubbles by describing the
asset price trajectory as a super-exponential power law acceleration
decorated with log-periodic oscillations, reflecting nonlinear positive feedbacks and herding effects among investors. 

In its original form, the residuals of the LPPL model were typically treated as white noise without a specific economic structure. A major
advance was later proposed by Lin, Ren, and Sornette (2014)\cite{Lin2014}, who introduced the so-called
‘volatility-confined LPPL model’. In this specification, the
residuals are modeled as an Ornstein--Uhlenbeck (O--U) mean-reverting
process, ensuring that deviations from the deterministic LPPL  trajectory remain bounded. This modification yields a consistent and
self-contained framework in which the deterministic LPPL component
represents the bubble dynamics, while the residuals account for the
stochastic reassessment of returns by investors.

% The study of financial bubbles has been substantially advanced by
% Robert A.~Jarrow and coauthors, who developed rigorous frameworks to
% characterize bubbles within modern asset pricing theory. In particular,
Jarrow, Protter, and Shimbo (2010)\cite{
JarrowProtterShimbo2010} established the martingale approach to bubbles
in both complete and incomplete markets, showing how strict local
martingales can represent bubble dynamics and affect derivative pricing.

For asset bubbles, Phillips, Shi, and Yu (2015)\cite{phillips2015testing} introduced the Generalized Sup ADF (GSADF) test to detect explosive behaviors in asset price time series. Their empirical analysis of the S\&P 500 revealed that the GSADF test statistic typically begins to diverge significantly about 2 to 3 months before a bubble collapses. This suggests that financial bubbles often exhibit measurable warning signs well before the actual crash.

\subsection{LPPL Model}
%  and our research foucs

\hspace{1.5em}This section serves to situate our research within the broader literature on bubble modeling. 
We begin by reviewing the Log-Periodic Power Law (LPPL) model developed by Sornette and collaborators, which provides the dominant theoretical framework for characterizing speculative bubbles as faster-than-exponential trajectories decorated with log-periodic oscillations. 
We then highlight the gap in this stream of research: while LPPL and its extensions (such as LPPLS) focus on technical price dynamics and the prediction of critical times, they tend to underemphasize behavioral-finance drivers such as investor sentiment and media attention. 
Finally, we introduce our own research focus, which aims to complement the LPPL tradition by systematically integrating behavioral signals with residual-based LPPL analysis, thereby transforming bubble detection into a richer and more operational framework.  

\subsubsection{LPPL model}
\hspace{1.5em}The Log-Periodic Power Law (LPPL) model, pioneered by Sornette and Johansen, represents a cornerstone in the quantitative analysis of speculative bubbles\cite{Sornette2017}. The LPPL framework originates from the analogy between financial markets approaching a crash and critical phase transitions in physical systems, where collective behavior leads to finite-time singularities. 

Formally, the LPPL model postulates that the logarithm of the asset price $p(t)$ can be expressed as

\begin{equation}
\ln p(t) = A + B (t_c - t)^m + C (t_c - t)^m \cos\!\Big(\omega \ln(t_c - t) + \phi \Big),
\label{eq:lppl}
\end{equation}

where $A$ is the constant baseline level, $B$ captures the super-exponential growth rate, $C$ determines the amplitude of oscillations, $t_c$ is the critical time (the theoretical end of the bubble), $m \in (0,1)$ is the critical exponent, $\omega > 0$ is the log-periodic frequency, and $\phi$ is the phase.

\vspace{0.5em}
\noindent
The equation above can be motivated in two steps:

1. Power-law acceleration.
Suppose that the expected growth of prices follows a nonlinear feedback mechanism such that
\begin{equation}
\frac{d \ln p(t)}{dt} \propto (t_c - t)^{m-1}, \qquad 0 < m < 1,
\end{equation}
which integrates to a finite-time singularity of the form
\begin{equation}
\ln p(t) = A + B (t_c - t)^m.
\end{equation}

2. Discrete scale invariance.
Empirical evidence shows oscillatory corrections to the power law. By assuming that investor herding generates log-periodic fluctuations, the growth term is decorated with an oscillatory modulation leading to (1).

\vspace{0.5em}
\noindent
The power-law component $(t_c - t)^m$ encodes the deterministic acceleration of prices as the bubble approaches its critical end point.  
The log-periodic term reflects alternating phases of optimism and skepticism among market participants, whose frequency increases as $t \to t_c$, signaling the growing instability of the bubble regime.  
Together, these terms imply that bubbles are not purely exponential but exhibit faster-than-exponential growth punctuated by accelerating oscillations.

This formulation has been widely applied in empirical studies of equity, real estate, and commodity bubbles, providing a theoretical backbone for the quantitative detection of financial instabilities.

\subsubsection{Limitations of LPPL Model}

\hspace*{2em}Despite significant progress in LPPL methodology and the integration of behavioral finance, several core limitations remain that constrain both practical applicability and theoretical completeness. First, conventional LPPL models primarily emphasize the detection of overvaluation bubbles, while neglecting undervaluation episodes that may represent equally important trading opportunities. Such an asymmetric perspective overlooks the fact that markets can systematically deviate from fair value in both directions. Second, most existing detection frameworks rely almost exclusively on price-based technical indicators, while disregarding the information contained in investor sentiment and market attention measures—factors that behavioral finance theory identifies as central to bubble dynamics. Third, current approaches often lack the dynamic adaptability necessary for real-time trading, producing retrospective assessments rather than actionable forward-looking signals.

Following Sornette's foundational work, subsequent research has introduced the Log-Periodic Power Law Singularity (LPPLS) model to enhance bubble detection accuracy and improve predictive capabilities. The LPPLS framework incorporates additional parameters to capture the complex oscillatory behavior preceding market crashes, representing a significant methodological advancement in technical bubble detection. However, while these developments have strengthened the mathematical rigor of price-based analysis, they maintain the fundamental limitation of focusing primarily on technical price patterns while neglecting the behavioral and psychological drivers that underlie bubble formation and evolution.

Our research diverges from this purely technical trajectory by prioritizing the systematic integration of behavioral finance principles into quantitative bubble detection frameworks. Rather than solely refining mathematical models of price dynamics, we focus on incorporating the human elements—investor sentiment, media attention, and market psychology—that behavioral finance theory identifies as the fundamental drivers of bubble phenomena. This approach recognizes that bubbles are fundamentally behavioral phenomena that manifest in price patterns, rather than purely mathematical anomalies that can be captured through technical analysis alone.

This study addresses these limitations through a set of theoretical and methodological contributions that advance the state of bubble detection. Our work spans four dimensions: theoretical framework development, methodological design, empirical validation, and practical implementation.

\subsubsection{Complementary Research Focus}

\hspace*{1.5em}The Log-Periodic Power Law (LPPL) and its extensions have provided a mathematically rigorous framework for modeling speculative bubbles and their associated crashes. Sornette’s pioneering work has demonstrated how super-exponential trajectories decorated with log-periodic oscillations can capture the collective dynamics of financial markets, leading to the identification of both bubble phases and their mirror images, such as negative bubbles and anti-bubbles. These contributions have established the LPPL family of models as a cornerstone in the quantitative analysis of market instability.

Our research does not aim to challenge or replace this stream of work. Rather, it focuses on a complementary dimension: the behavioral layer that underpins bubble phenomena. While LPPL emphasizes the endogenous dynamics of price trajectories around critical times, our framework seeks to quantify how media attention, investor sentiment, and broader market psychology contribute to deviations from the LPPL-predicted path. In this sense, our study offers an additional behavioral-finance-based perspective that enriches the LPPL tradition.

Specifically, we are interested in identifying what we call negative behaviors—periods during an ongoing bubble trajectory when observed prices fall below the LPPL fit but do not evolve into full-fledged negative bubbles. These episodes often correspond to temporary corrections or oversold phases, which may present actionable trading opportunities. In many cases, such negative-bubble phases are followed by sharp or prolonged overbuy rebounds, pushing the bubble to new highs; indeed, a full bubble trajectory often only collapses after undergoing several rounds of alternating overselling and overbuying.  By embedding residual-based measures into a behavioral framework, we extend LPPL modeling toward a more comprehensive detection system that highlights both overpricing and underpricing dynamics.

In this sense, our research represents a bridge between technical modeling and behavioral finance. Whereas the LPPL family formalizes the mathematical structure of bubbles, our contribution lies in integrating sentiment analysis, hype indices, and other behavioral indicators into the detection process. This quantitative innovation not only complements existing LPPL-based methodologies but also moves closer to capturing the inherently behavioral nature of financial bubbles.

\subsection{NLP for Finance}

\hspace{1.5em}NLP-based news sentiment has proven effective in capturing shifts in market conditions. Shapiro, Sudhof, and Wilson (2017)\cite{shapiro2017measuring} develop an economic sentiment index based on real-time financial news, showing that sentiment predicts macro-variables such as consumption and output. Similarly, Tetlock (2007)\cite{tetlock2007giving} constructs a media-based sentiment index from daily Wall Street Journal columns, finding that negative news sentiment significantly predicts lower market returns. These studies highlight how unstructured news content embeds investor sentiment and provides predictive power for asset prices.

Building on this line of research, Maghyereh and Abdoh (2022)\cite{maghyereh2022news} examine whether news-based economic sentiment can predict price bubbles in precious metals markets. Using GSADF tests and probit models, they find that bearish sentiment significantly raises the probability of bubbles in gold and platinum, particularly during crisis periods. This result underscores the predictive value of sentiment signals relative to traditional market indicators.

% \subsection{Hype Index} \label{subsec:hype_index}
% \hspace{1.5em}
The role of media attention in asset pricing has long been recognized in financial economics, with early work by Tetlock (2007) demonstrating how media pessimism predicts short-term market returns and trading activity. This line of research was advanced by Glasserman and Mamaysky (2019)\cite{glasserman2019}, who introduced measures of informational ``unusualness'' in financial news and linked them to future volatility. Cao and Geman (2025) have shown that using improved sentiment analysis and scoring methods and the change of probability measure can greatly enhance the forecasting accuracy of equity price and volatility directions\cite{caogeman2025}.

Within this broader literature, the Hype Index was proposed by Cao, Wunkaew, and Geman (2025)\cite{caowunkaewgeman2025} as a novel, NLP-driven measure of investor attention. Unlike traditional sentiment indicators, the Hype Index isolates the intensity of media coverage rather than its tone. By quantifying the share of financial news devoted to a given stock or sector, it provides a direct measure of disproportionate attention that is distinct from sentiment polarity\cite{caogeman2025}. The authors further extend the concept through a capitalization-adjusted version, which benchmarks media exposure relative to economic size, thereby identifying firms or sectors that attract excess attention beyond what fundamentals would suggest.  

Empirical evaluation shows that the Hype Index is systematically associated with volatility, market-wide stress events, and cross-sector differences in attention. For instance, sectors such as Information Technology and Financials are persistently over-hyped, while Utilities and Real Estate remain under-hyped over long horizons. This framework thus broadens the toolkit for analyzing bubbles, volatility clustering, and behavioral biases, providing an interpretable bridge between unstructured news flows and market dynamics.

\vspace{1.2em}
\section{The Hyped Log-Periodic Power Law Model} \label{section:BubbleScore}

\hspace{1.5em}This section develops the conceptual and methodological foundation of the proposed \emph{{Bubble Score}}. We integrate financial theory, behavioral measures, and the Log-Periodic Power Law (LPPL) framework to capture both speculative bubbles and protracted undervaluation phases. The {Bubble Score} is designed as a unified indicator, capable of distinguishing overvaluation (bubble behaviors) from undervaluation (negative behaviors), while embedding market psychology through hype and sentiment. 

\subsection{Financial Bubbles and Negative Bubbles}

\hspace{1.5em}Financial bubble is generally understood as a phase of accelerating overvaluation in which asset prices grow at a faster-than-exponential rate. This growth is fueled by positive feedback loops among investors, herding behavior, and speculative enthusiasm. Prices are pushed far above levels justified by fundamentals, and the process culminates in a regime shift, often marked by a sharp correction or crash. Within the Log-Periodic Power Law (LPPL) framework of Sornette, bubbles are modeled as follows,

\begin{equation}
\ln p(t) = A + B (t_c - t)^m + C (t_c - t)^m \cos\!\Big(\omega \ln(t_c - t) + \phi \Big),
\end{equation}

where $t_c$ is the critical time, $m \in (0,1)$, and the oscillatory term reflects alternating waves of optimism and skepticism. As $t \to t_c$, the oscillations compress in time, indicating a growing instability in the system.  

a long-lived phase of protracted decline following a crash, modeled in the LPPL framework by reversing the temporal symmetry of the bubble trajectory. However, in this study we do not explicitly model such full-fledged negative bubbles. Instead, our focus remains on the bubble trajectory, while recognizing that prices may temporarily fall below the fitted LPPL path.  

We refer to these deviations as negative behaviors: phases in which observed prices lie persistently below the LPPL trajectory, without constituting a structural negative bubble. Negative behaviors are of particular interest because they may signal oversold conditions within an ongoing bubble regime, thereby offering potential buy-in opportunities before the bubble resumes or intensifies. This terminology allows us to capture bearish mispricings in a bubble context, while keeping the analytical framework firmly grounded in the dynamics of the LPPL bubble model.

\subsection{LPPL Residuals and Normalization}

To operationalize deviations from the LPPL benchmark, we define the log-price residual as
\begin{equation}
\epsilon(t) = \ln p(t) - \ln \hat{p}(t),
\end{equation}
where $p(t)$ is the observed price and $\ln \hat{p}(t)$ is the LPPL trajectory as shown in (5).

Within the extended study of Lin, Ren, and Sornette (2014), 
these residuals are not treated as random walks, but as mean-reverting 
fluctuations around the LPPL path. 

Formally, the dynamics of residuals $\epsilon(t)$ 
can be expressed as
\begin{equation}
\Delta \epsilon(t) = \epsilon(t+1) - \epsilon(t)= -\alpha \epsilon(t) + u_t,
\end{equation}
where $u_t$ is a Gaussian white noise. This AR(1) specification ensures that 
$\epsilon(t)$ remains stationary, consistently pulling the log-price back towards the 
LPPL trajectory.

Equivalently, the log-price process satisfies
\begin{equation}
\ln p_{t+1}  = \ln p_t + \Delta \ln \hat{p}(t) - \alpha \big(\ln p_t - \ln \hat{p}(t)\big) + u_t,
\end{equation}
so that the observed trajectory is composed of a deterministic LPPL drift plus 
a mean-reverting residual. This guarantees that explosive LPPL growth is 
"confined" by volatility rather than diverging arbitrarily.

For comparability across assets and time windows, we normalize residuals by 
their maximum absolute deviation:
\begin{equation}
\epsilon_{\text{norm}}(t) = \frac{\epsilon(t)}{\max_{s \leq t} |\epsilon(s)|}, 
\qquad \epsilon_{\text{norm}}(t) \in [-1,1].
\end{equation}

Positive values of $\epsilon_{\text{norm}}(t)$ are interpreted as bubble behaviors, where observed prices stand above the LPPL benchmark and exhibit faster-than-exponential growth.  
Negative values of $\epsilon_{\text{norm}}(t)$ instead signal negative behaviors, corresponding to phases of relative underpricing or incomplete corrections in which prices persist below the LPPL trajectory.

Thus, the residual process provides a mathematically consistent way to capture 
short-term mispricings while preserving the global LPPL bubble dynamics.

\subsection{Sentiment Score} \label{subsec:sentiment_score}

\hspace{1.5em}While the Hype Index measures the intensity of media coverage, it does not account for the tone of that coverage. To complement it, we incorporate a Sentiment Score that captures the polarity of news content. This follows a long line of empirical work, beginning with Tetlock (2007), which demonstrates that media pessimism predicts short-term returns and trading volume, and more recent advances in NLP-based sentiment analysis.

\textbf{Definition.} Let each news article $k$ about stock $i$ on day $t$ be assigned a sentiment polarity score $s_{i,t,k} \in [-1,1]$, where $-1$ denotes extreme pessimism, $0$ neutrality, and $+1$ extreme optimism. Using the Finbert sentiment model, these scores are computed based on lexical and syntactic features of the text. The aggregate Sentiment Score for stock $i$ on day $t$ is then defined as the weighted average:
\begin{equation}
S_{i,t} = \frac{\sum_{k=1}^{N_{i,t}} w_{i,t,k}\, s_{i,t,k}}{\sum_{k=1}^{N_{i,t}} w_{i,t,k}},
\label{eq:sentiment_score}
\end{equation}
where $N_{i,t}$ is the number of articles about stock $i$ on day $t$, and $w_{i,t,k}$ is a weight that can reflect article length, source credibility, or recency.  
% Vader

The Sentiment Score is scale-free, bounded in $[-1,1]$, and directly comparable across time and assets. It embeds the psychological tone of news flow into the analytical framework, thereby complementing the Hype Index. Together, $(H_{i,t}, S_{i,t})$ provide a joint characterization of media attention and tone, enabling richer detection of bubble-related behaviors than price-based signals alone.

\subsection{Hype Index} \label{subsec:hype_index}

The role of media attention in asset pricing has been repeatedly highlighted in the literature. Investor attention is limited, and disproportionate coverage of certain assets creates systematic patterns in both returns and volatility. Building on this idea, Cao, Wunkaew, and Geman (2025) introduce the Hype Index as a quantitative measure of the relative share of media coverage devoted to a given stock or sector. Unlike sentiment indicators that focus on the polarity of news, the Hype Index isolates the intensity of coverage, thereby disentangling how much attention an asset receives from how that attention is framed.  

\textbf{Definition.} Let $N_{i,t}$ denote the number of financial news articles that mention stock $i$ on day $t$, and let
\begin{equation}
N_{\text{mkt},t} = \sum_{j=1}^{M} N_{j,t}
\end{equation}
be the aggregate number of articles covering a reference set of $M$ firms (e.g., the S\&P 100 constituents) on the same day. The \emph{Hype Index} for stock $i$ is defined as
\begin{equation}
H_{i,t} = \frac{N_{i,t}}{N_{\text{mkt},t}}, 
\qquad \sum_{i=1}^{M} H_{i,t} = 1.
\end{equation}
This construction makes $H_{i,t}$ a probability measure of media attention allocation across the universe of reference stocks.  

A higher value of $H_{i,t}$ indicates disproportionate media focus relative to peers, regardless of the firm’s market capitalization or fundamentals. When $H_{i,t}$ rises sharply, the stock dominates financial news flow, capturing a larger share of scarce investor attention. This aligns with attention-based asset pricing theory, which posits that such concentration of focus can drive both mispricing and volatility clustering.  

To differentiate between “natural” attention due to economic importance and “excessive” attention, Cao et al. (2025) also introduce a capitalization-adjusted version:
\begin{equation}
\mathrm{CapH}_{i,t} = \frac{H_{i,t}}{W^{\text{cap}}_{i,t}}, 
\qquad W^{\text{cap}}_{i,t} = \frac{MC_{i,t}}{\sum_{j=1}^{M} MC_{j,t}},
\end{equation}
where $MC_{i,t}$ is the market capitalization of firm $i$ at time $t$.  
The capitalization-adjusted Hype Index $\mathrm{CapH}_{i,t}$ can be interpreted as
\begin{equation}
\mathrm{CapH}_{i,t} =
\begin{cases}
> 1, & \text{excessive hype: media coverage exceeds the stock’s economic weight}, \\
< 1, & \text{under-hype: media coverage falls short of the level implied by capitalization}.
\end{cases}
\end{equation}

This dual representation allows us to separate absolute media attention from relative overexposure. Persistent deviations of $\mathrm{CapH}_{i,t}$ above unity are indicative of hype-driven dynamics that can reinforce bubble behaviors, while values below unity suggest neglected or under-covered stocks. The Hype Index thus provides a mathematically grounded, scale-invariant way to quantify the role of media in amplifying both bubble and negative behavior regimes.

\subsection{Bubble Score}

\hspace{1.5em}We combine technical mispricing with behavioral signals into a single Bubble Score:
\begin{equation}
\text{BubbleScore}_{i}(t) =
\begin{cases}
\epsilon_{\text{norm}}(t) + \alpha_1 H_{i,t} + \alpha_2 S_{i,t}, & \epsilon_{\text{norm}}(t) > 0 \ (\text{Bubble behavior}), \\[6pt]
\epsilon_{\text{norm}}(t) - \alpha_1 H_{i,t} + \alpha_2 S_{i,t}, & \epsilon_{\text{norm}}(t) < 0 \ (\text{Negative behavior}).
\end{cases}
\end{equation}
Here, $\alpha_1$ and $\alpha_2$ are  parameters that determine the relative weight of media attention and sentiment. The construction reflects the intuition that hype always amplifies speculative intensity, while sentiment exerts opposite effects in bubble versus negative behavior regimes.  

{
The construction in equation (13) reflects different roles of hype and sentiment. 
Hype $H_{i,t}$ amplifies market extremes: it is positive in bubble phases to reinforce overvaluation, 
and negative in negative-bubble phases to deepen undervaluation. 
Sentiment $S_{i,t}$ acts as a corrective buffer: in bubble phases, optimistic sentiment sustains the upward momentum 
while pessimistic sentiment dampens it; in negative-bubble phases, optimistic sentiment alleviates undervaluation 
while pessimistic sentiment intensifies it. 

Overall, the BubbleScore measures the likelihood and intensity of a bubble regime. 
Large positive values indicate strong speculative overpricing (normal bubbles), 
while large negative values indicate pronounced undervaluation negative bubbles). 
The magnitude of the score thus provides a direct quantitative gauge of bubble dynamics.
}

Positive values indicate speculative overvaluation, while negative values indicate underpricing relative to LPPL expectations. Magnitudes reflect the intensity of deviation.

\subsection{Case Study: Across Different Individual Stocks in U.S. Market}

\hspace{1.5em}Before turning to specific examples, we illustrate how the Bubble Score developed in the previous section can be applied as an indicator for empirical analysis. In particular, when the Bubble Score remains above a chosen threshold for a sustained period, we classify the corresponding interval as a normal bubble period, whereas periods of sustained negative values are designated as negative bubble periods. This classification enables us to visualize bubble dynamics directly on daily price trajectories. Otherwise, we call the trajectory during a bubble neutral period.

\subsubsection{HOUS Case Study}

\begin{figure}[H]
    \centering
    \includegraphics[width=0.95\textwidth]{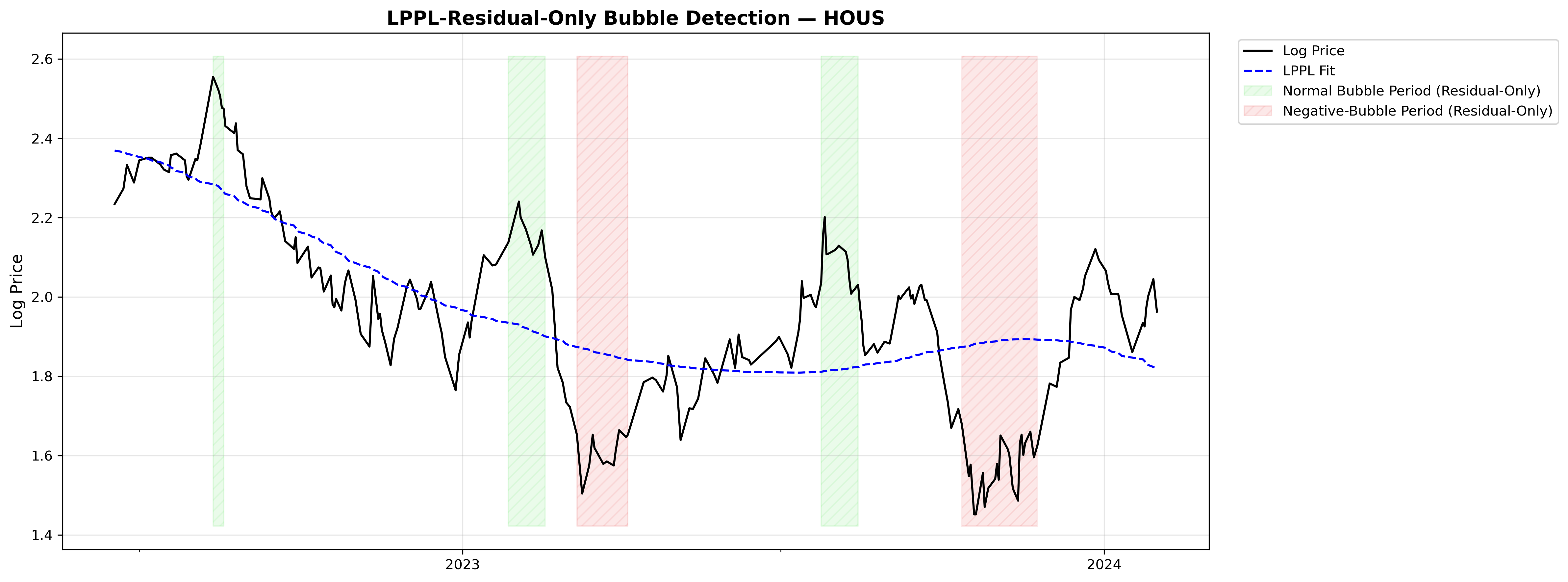}
    \caption{LPPL \emph{residual-only} detection for HOUS. The black line shows log prices and the blue dashed curve the LPPL fit. Green (red) bands denote intervals where the observed series is materially above (below) the theoretical path. A clear local peak is detected in mid-2023, followed by a prolonged oversold phase into late-2023; however, short and noisy swings can still trigger event windows under the residual-only approach.}
    \label{fig:hous_bubble_detection}
\end{figure}

\begin{figure}[H]
    \centering
    \includegraphics[width=0.95\linewidth]{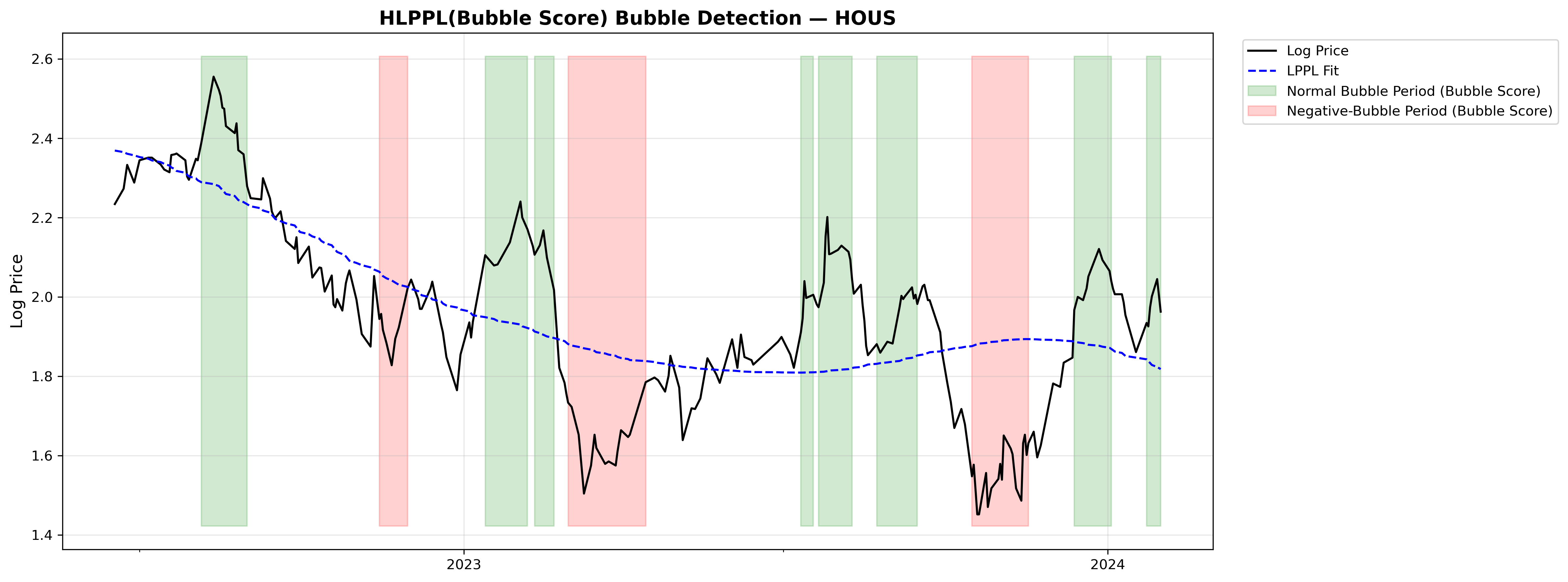}
    \caption{HLPPL (BubbleScore) detection for HOUS. By augmenting residual information with a Hype index and text Sentiment, and enforcing minimum-duration and threshold rules, spurious short-lived windows are filtered out and the alternating bubble/negative-bubble phases over 2022–2024 are delineated more crisply (including the mid-2023 top and the late-2023 to early-2024 rebound).}
    \label{fig:hous_bubblescore_detection}
\end{figure}

\begin{figure}[H]
    \centering
    \includegraphics[width=0.95\linewidth]{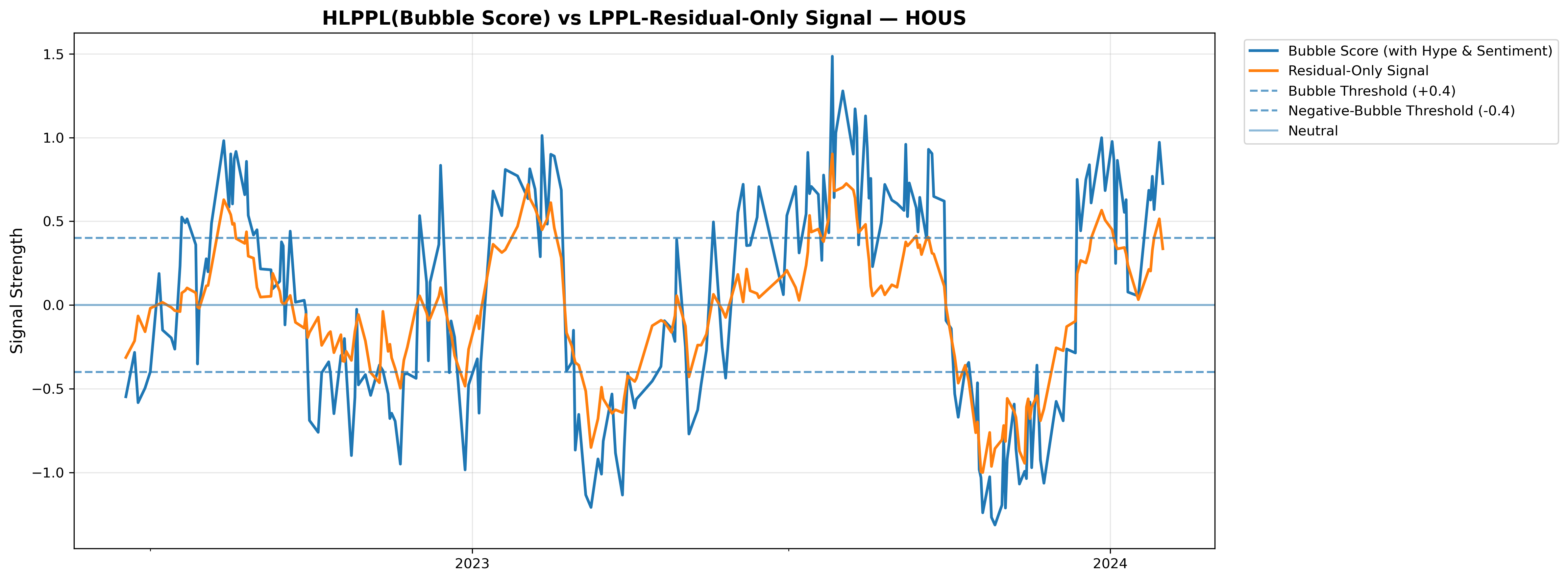}
    \caption{Signal comparison: BubbleScore (with Hype \& Sentiment) vs.\ residual-only. The two signals agree on major turning points, but HLPPL is smoother around decision thresholds (e.g., $\pm 0.4$), produces fewer flips during choppy regimes, and down-weights moves that lack media attention or textual-emotion corroboration—reducing false positives/negatives and improving tradeability.}
    \label{fig:hous_signal_compare}
\end{figure}

\begin{figure}[H]
    \centering
    \includegraphics[width=0.95\linewidth]{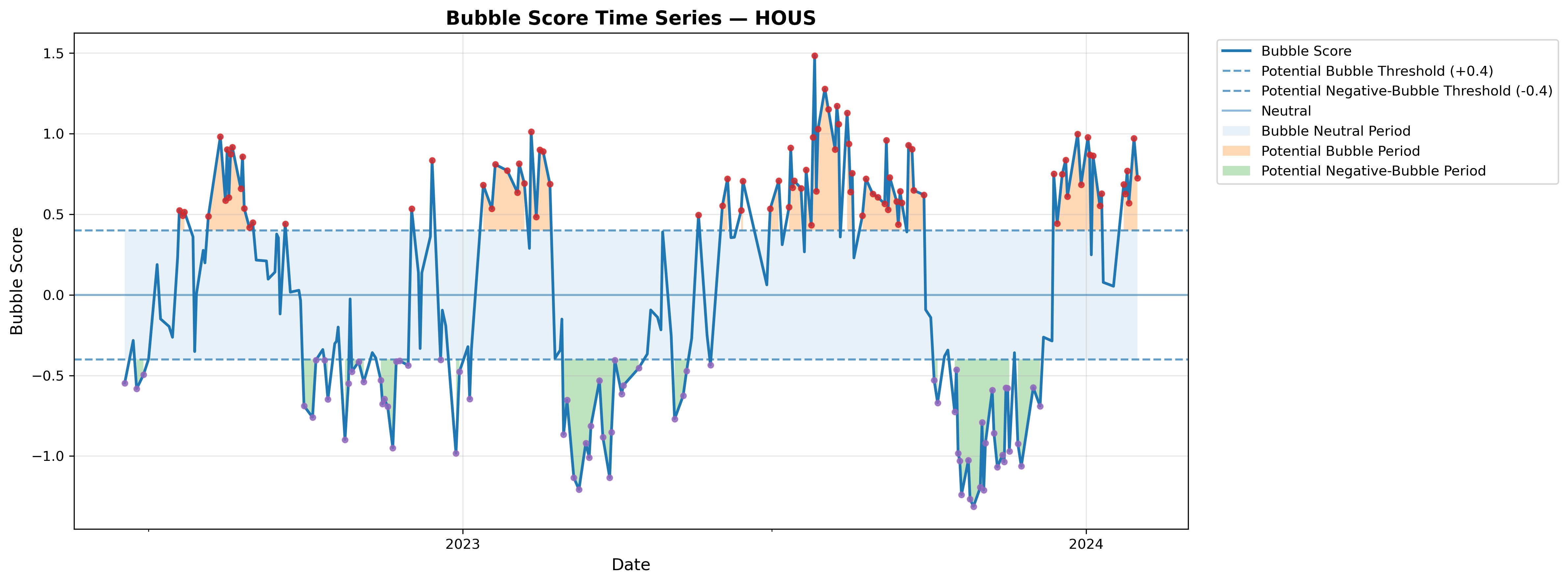}
    \caption{Daily BubbleScore time series for HOUS with threshold bands and event shading. The sequence of up-crossings in mid-2023, subsequent down-crossings in late-2023, and re-entries above the upper band in early-2024 illustrates the regime rhythm under a unified ``behavioral + technical'' lens. Peak markers assist rule-based entries, scaling, and profit-taking.}
    \label{fig:hous_bubblescore_ts}
\end{figure}

\noindent
Figure~\ref{fig:hous_bubble_detection}–\ref{fig:hous_bubblescore_ts} apply our BubbleScore framework to Anywhere Real Estate Inc.\ (HOUS), a leading U.S.\ residential real-estate brokerage operating brands such as Century 21 and Coldwell Banker. Because our later trading experiments target housing-related exposures, HOUS is a natural benchmark to validate both relevance and robustness of our methodology in this sector.

For HOUS, HLPPL (BubbleScore) systematically outperforms a residual-only specification by delivering more robust and tradeable signals. By jointly weighting LPPL residuals with a Hype index and text-based Sentiment, the measure suppresses idiosyncratic fluctuations that typically arise in sideways regimes, thereby lowering false discoveries near decision thresholds. Under a common set of thresholds and minimum-duration constraints, the resulting event windows exhibit sharper temporal boundaries and more consistent lengths, which facilitates rule-based entries, scaling, and exits. Moreover, the framework naturally reconciles technical and behavioral information: when residuals and text indicators align, the score concentrates mass in genuine extremes; when they diverge, the contribution of price-only noise is down-weighted. In aggregate, BubbleScore preserves the sensitivity of LPPL to critical dynamics while producing steadier, operationally actionable signals for housing-exposed strategies.

\subsubsection{AMTX Case Study}

\noindent
Aemetis Inc.\ (AMTX) provides another representative case study for our framework. It is a U.S.-based renewable fuels and biochemicals company founded in 2006 and headquartered in Cupertino, California. It was selected here as an illustrative example due to its unusually large price swings in recent years. From 2018 to 2025, the stock experienced multiple sharp rallies and corrections, making it a natural testbed for assessing the robustness of our BubbleScore methodology on daily data.

\begin{figure}[H]
    \centering
    \includegraphics[width=0.95\textwidth]{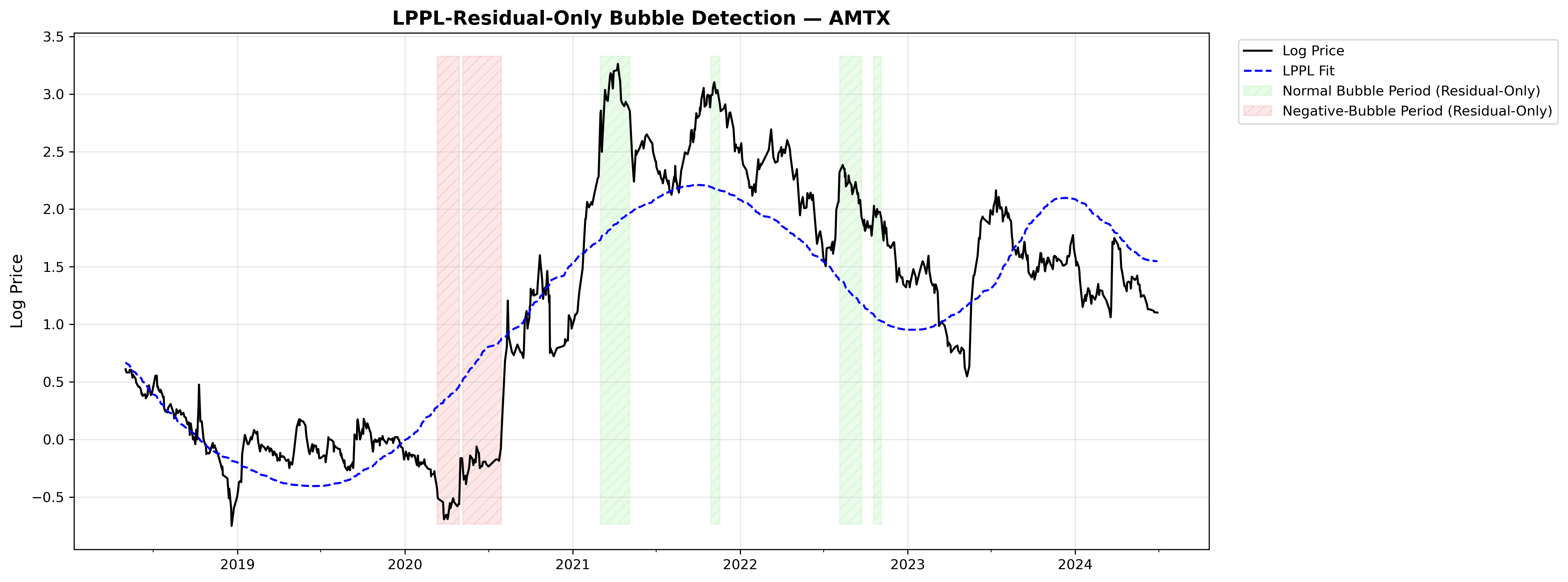}
    \caption{LPPL \emph{residual-only} detection for AMTX. The black line shows the log price and the blue dashed curve the LPPL fit. Green (red) shading marks intervals where prices lie materially above (below) the theoretical path. Distinct positive-bubble episodes appear in 2021–2022, while early-2018 and mid-2020 exhibit pronounced negative-bubble behavior.}
    \label{fig:amtx_bubble_detection}
\end{figure}

\begin{figure}[H]
    \centering
    \includegraphics[width=0.95\linewidth]{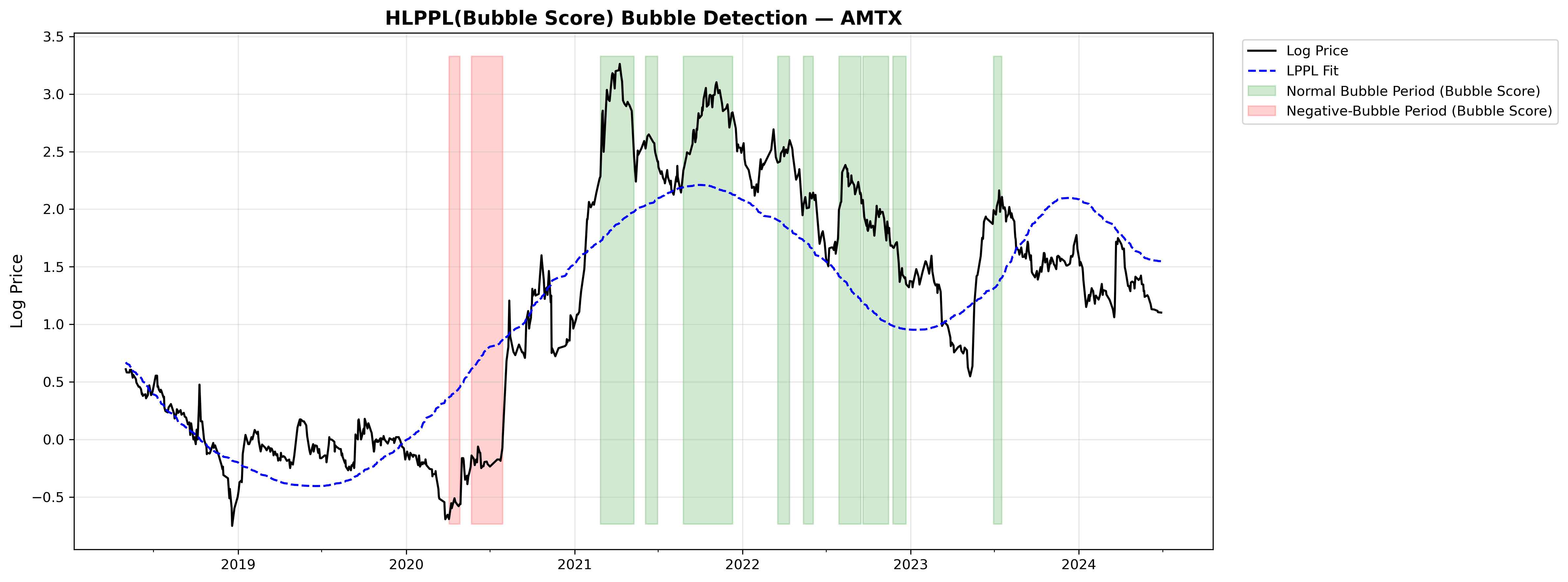}
    \caption{HLPPL (BubbleScore) detection for AMTX. Relative to residual-only tagging, augmenting with Hype and text-based Sentiment, together with minimum-duration and threshold rules, filters out short-lived spikes and yields cleaner, temporally coherent windows across 2018–2025.}
    \label{fig:amtx_bubblescore_detection}
\end{figure}

\begin{figure}[H]
    \centering
    \includegraphics[width=0.95\linewidth]{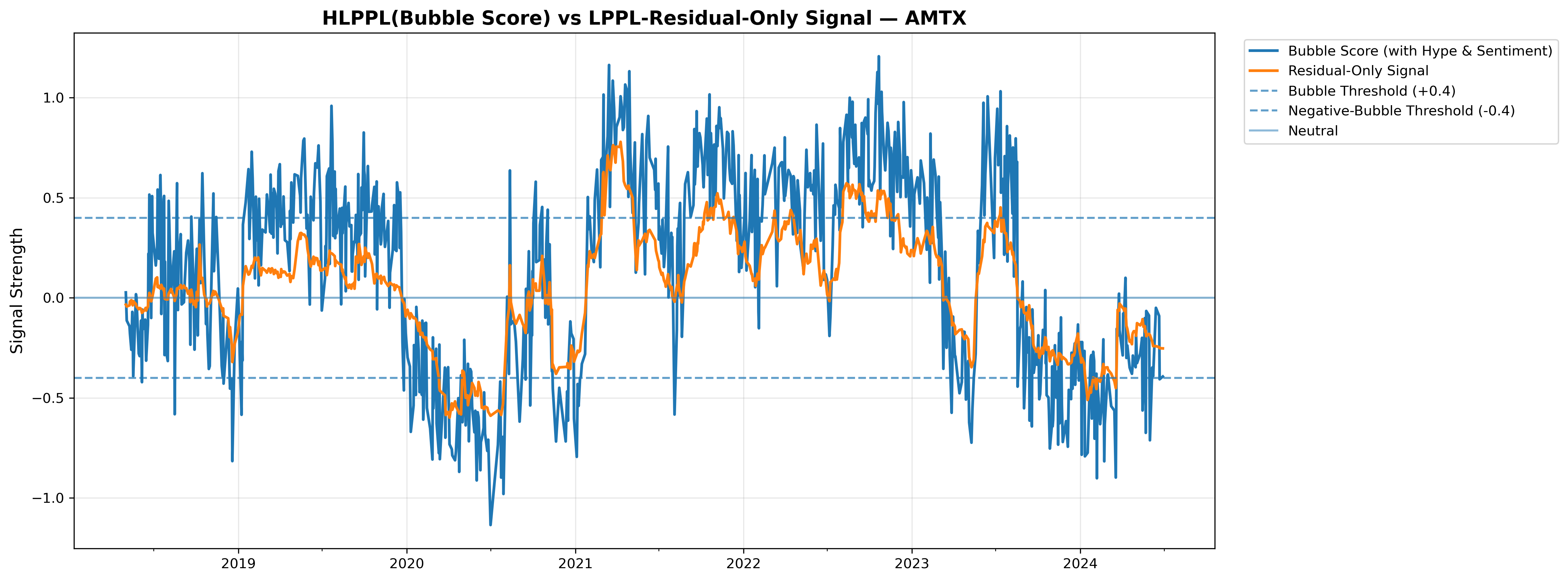}
    \caption{Signal comparison for AMTX: BubbleScore (with Hype \& Sentiment) versus residual-only. The series agree on major turning points, but HLPPL is smoother around decision bands (e.g., $\pm 0.4$), exhibits fewer flips in volatile stretches, and down-weights price moves that lack behavioral corroboration—reducing false positives/negatives.}
    \label{fig:amtx_signal_compare}
\end{figure}

\begin{figure}[H]
    \centering
    \includegraphics[width=0.95\linewidth]{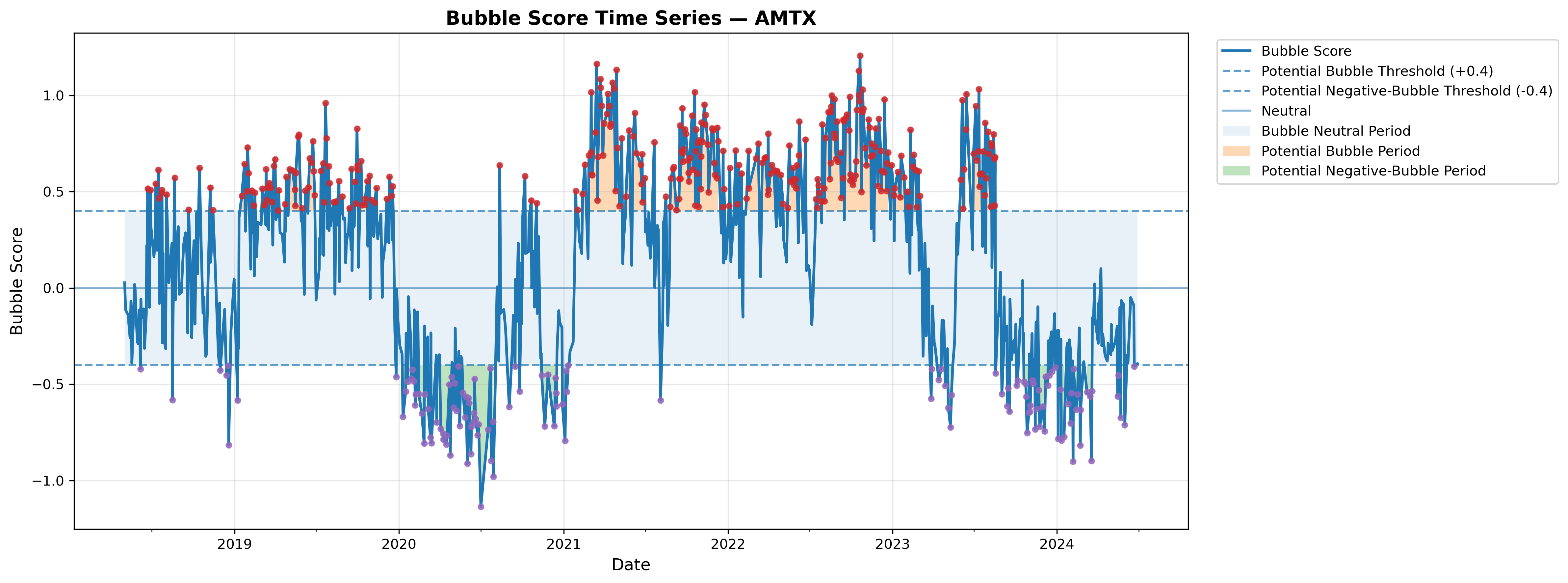}
    \caption{Daily BubbleScore time series for AMTX with threshold bands and shaded event segments. Peak markers indicate local extremes used for rule-based entries, scaling, and exits; the sequence of up-/down-crossings from 2018 to 2025 reveals a consistent regime structure despite high volatility.}
    \label{fig:amtx_bubblescore_ts}
\end{figure}

 AMTX’s pronounced volatility motivates the use of a longer minimum-duration parameter (\texttt{min\_days}) to prevent short, high-amplitude swings from triggering spurious events. Even under this stricter setting, the framework isolates the core bubble dynamics: BubbleScore identifies the 2021–2022 surges as sustained positive-bubble phases and cleanly separates earlier negative-bubble intervals (early-2018, mid-2020). By jointly weighting LPPL residuals with Hype and Sentiment, BubbleScore preserves sensitivity to LPPL critical dynamics while attenuating noise in choppy markets, delivering sharper window boundaries, fewer threshold flips, and more operationally actionable signals on daily data.

\subsection{Beyond Technical and Behavioral Indicators}

\hspace{1.5em}While the {Bubble Score} integrates LPPL residuals, hype, and sentiment, we acknowledge that certain cases challenge its interpretation. For example, firms with extraordinary fundamentals (e.g., Nvidia between 2018--2022) may exhibit super-exponential price paths that are justified by earnings and innovation rather than speculation. Detecting such “fundamentally supported growth” requires information beyond price dynamics and textual signals.  

To address this limitation, we extend the framework in the next section by incorporating machine learning models that combine the {Bubble Score} with additional market and fundamental features (e.g., valuation ratios, macro indicators). This integration aims to refine detection accuracy, reduce false positives, and identify genuine bubbles versus justified growth trajectories.

\section{Transformer Model} \label{section: transformer model}

\hspace{1.5em}We develop a supervised learning framework for asset-level bubble detection by integrating traditional econometric tests with deep neural sequence modeling. This hybrid approach captures both statistically-grounded bubble signals and complex temporal patterns embedded in financial time series. The objective of this model is to forecast the bubble intensity ({Bubble Score}) over the next five trading days.

\subsection{Model Structure and Training Loss}

\hspace{1.5em}We begin by constructing a comprehensive set of input features reflecting macroeconomic sentiment, valuation ratios, and market behavior. These include variables such as the Volatility Index (VIX), Hype Index, news sentiment scores (from FinBERT), price-to-earnings (PE) and price-to-book (PB) ratios, as well as raw market signals like closing prices and trading volume. These inputs collectively capture fundamental value, investor psychology, and liquidity dynamics.

Next, the raw time series are normalized and transformed into structured feature matrices to encode temporal dependencies. Each matrix represents a sequence of past values, preserving historical context for downstream modeling.

To fully exploit the heterogeneous nature of the inputs—specifically the coexistence of asset-level features and market-level signals—we employ a Dual-Stream Transformer architecture. One Transformer encoder models stock-specific sequences, while another processes market-wide signals. The dual-stream Transformer framework can jointly models stock-level and market-level information. The input consists of two parallel sequences
\[
X^{(s)}_{1:t} = \{x^{(s)}_1,\ldots,x^{(s)}_t\}, \quad 
X^{(m)}_{1:t} = \{x^{(m)}_1,\ldots,x^{(m)}_t\}, \quad 
x^{(\cdot)}_s \in \mathbb{R}^d ,
\]
where $x^{(s)}_s$ denotes asset-specific features and $x^{(m)}_s$ represents market-wide signals.

Each input is linearly projected and enriched with positional information:
\[
Z^{(s)} = \mathrm{PE}(W^{(s)}_e X^{(s)}), \qquad
Z^{(m)} = \mathrm{PE}(W^{(m)}_e X^{(m)}).
\]

The two sequences are processed by separate Transformer encoders:
\[
H^{(s)} = \mathrm{Enc}^{(s)}(Z^{(s)}), \qquad 
H^{(m)} = \mathrm{Enc}^{(m)}(Z^{(m)}).
\]

A bi-directional cross-attention mechanism is applied after the independent 
encoders to capture interactions between stock-level and market-level 
representations. This allows each stream not only to preserve its own temporal 
dynamics but also to attend to complementary information from the other, 
thereby integrating local asset behavior with broader market context.

Formally, given encoded sequences $H^{(s)}$ and $H^{(m)}$, we compute
\[
\widetilde{H}^{(s)} = \mathrm{Attn}(H^{(s)}W_Q^{(s)}, \, H^{(m)}W_K^{(m)}, \, H^{(m)}W_V^{(m)}), \qquad
\widetilde{H}^{(m)} = \mathrm{Attn}(H^{(m)}W_Q^{(m)}, \, H^{(s)}W_K^{(s)}, \, H^{(s)}W_V^{(s)}).
\]

The cross-attended representations are then pooled and fused into a joint 
feature vector,
\[
h_t = \phi\!\big(\mathrm{Pool}(\widetilde{H}^{(s)}), \, \mathrm{Pool}(\widetilde{H}^{(m)})\big),
\]
which serves as the common input for multiple prediction heads that forecast 
bubble intensity over the next five trading days:
\[
\hat{y}_{t+\tau} = f_\tau(h_t), \qquad \tau = 1,\ldots,5.
\]

Model parameters are optimized by minimizing an improved loss function that goes beyond a simple mean squared error and incorporates additional regularization objectives to better capture bubble dynamics. The detailed formulation of this loss will be introduced in a later section.

The architecture explicitly separates stock- and market-level signals, processing them through independent streams before fusing at a later stage. 
This design allows the model to capture both fine-grained local dynamics and broader market-wide context in financial time series. To train such a model effectively, we require supervision that not only identifies the presence of bubbles but also reflects their temporal evolution.

For this purpose, we employ the LPPL test to construct continuous labels. The LPPL framework yields a {Bubble Score} that quantifies the intensity of explosive dynamics in asset prices or trading 
volumes. Unlike binary labels, this index captures both bubble and reversal regimes, providing richer supervision and enabling the model to learn nuanced bubble dynamics.

Finally, the fused representations from the dual-stream Transformer are passed to multiple horizon-specific heads, each implemented as a lightweight multi-layer perceptron (MLP) with nonlinear activations and a Tanh output. The MLP serves as a flexible function approximator that maps the high-dimensional hidden representation into scalar {Bubble Score} forecasts, while its shallow architecture ensures computational efficiency and prevents overfitting. These heads separately forecast the {Bubble Score} for the next five trading 
days, allowing the model to capture horizon-specific dynamics while reducing conflicts in multi-step prediction.

\paragraph{Dataset Structure}
The dataset is chronologically partitioned into training, evaluation, and test sets in alignment with our three-stage training pipeline. The training set is used exclusively for optimizing model weights. The evaluation set is reserved for hyperparameter tuning, ensuring that parameter choices are not biased by the training data. The test set is employed for final performance assessment. This structured design allows a clear separation of roles—weight learning, hyperparameter selection, and unbiased testing—ensuring methodological rigor and reproducibility.

\paragraph{Regularization and Overfitting Prevention} 
To further enhance generalization under the noisy and volatile conditions of 
financial time series, we integrate multiple regularization strategies into 
the training process. 

First, dropout layers are applied in the input projection, feature fusion, 
and prediction heads to reduce neuron co-adaptation, formally:
\[
h = \mathrm{Dropout}(p)\big(f(x)\big), \qquad p=0.1\text{--}0.3,
\]
where $p$ denotes the dropout probability. 

Second, L2 weight decay is imposed on model parameters to control complexity:
\[
\mathcal{L}_{\text{reg}} = \lambda \| \theta \|_2^2,
\]
which penalizes excessively large weights. 

Third, gradient clipping is employed to stabilize updates and prevent exploding 
gradients:
\[
g \;\leftarrow\; g \cdot \min\!\Big(1, \tfrac{\tau}{\|g\|_2}\Big),
\]
where $g$ is the raw gradient and $\tau=0.5$ is the clipping threshold.

In addition, a OneCycleLR schedule dynamically adjusts the learning rate
\[
\eta_t = \eta_{\min} + \tfrac{1}{2}\big(\eta_{\max} - \eta_{\min}\big)
\Big(1 + \cos(\tfrac{\pi t}{T})\Big),
\]
ensuring smoother convergence. Finally, early stopping monitors validation loss 
and halts training once performance plateaus.

Together, these techniques mitigate overfitting and improve the robustness of the learned {Bubble Score} predictions.

\paragraph{Combined Loss Function.}
To better capture the complex dynamics of bubbles, we design a combined loss function that integrates multiple complementary objectives rather than relying on a single error metric. Specifically, the overall loss is formulated as a weighted sum of: (i) a Huber loss to provide robustness against outliers, (ii) a correlation loss to encourage alignment between predicted and true temporal patterns, (iii) an $R^2$-based loss to directly optimize model fit, (iv) a temporal consistency term to penalize discrepancies in day-to-day changes, and (v) a smoothness regularizer to discourage abrupt fluctuations in predictions. This multi-component formulation improves stability, enhances predictive accuracy, and ensures that the learned {Bubble Score} trajectories reflect both statistical fidelity and realistic temporal structure.

The overall training objective is defined as a weighted sum of multiple complementary terms:
\begin{equation}
\mathcal{L} = 
\lambda_{1}\,\mathcal{L}_{\text{Huber}}
+ \lambda_{2}\,\mathcal{L}_{\text{Corr}}
+ \lambda_{3}\,\mathcal{L}_{R^2}
+ \lambda_{4}\,\mathcal{L}_{\text{Cons}}
+ \lambda_{5}\,\mathcal{L}_{\text{Smooth}},
\end{equation}
where $\mathcal{L}_{\text{Huber}}$ denotes the Huber loss for robust regression, 
$\mathcal{L}_{\text{Corr}}$ enforces correlation between predicted and true sequences, 
$\mathcal{L}_{R^2}$ optimizes the coefficient of determination, 
$\mathcal{L}_{\text{Cons}}$ penalizes inconsistency in temporal differences, 
and $\mathcal{L}_{\text{Smooth}}$ regularizes abrupt changes. 

This formulation is motivated by the unique challenges of bubble prediction. Financial bubble dynamics are rare, noisy, and often exhibit abrupt changes, which makes single-objective loss functions insufficient. By combining robustness (Huber), statistical alignment (correlation and $R^2$), and temporal structure constraints (consistency and smoothness), the loss function guides the model to not only fit the data accurately but also generate predictions that are stable and economically plausible.

\begin{figure}[H]
\centering
\begin{tikzpicture}[
  node distance=1.4cm and 2.6cm,
  every node/.style={draw, align=center, rounded corners, minimum height=1.1cm, font=\small},
  arrow/.style={-{Latex}, thick}
]

% Stock path
\node (stock_input) {Stock-Level Features\\(Close Price, Volume,\\Other Variables)};
\node (stock_feature) [below=of stock_input] {Stock-Level Feature Matrix};

% Market path
\node (market_input) [right=of stock_input] {Market-Level Features\\(Hype Index, News Sentiment,\\Macro Indicators)};
\node (market_feature) [below=of market_input] {Market-Level Feature Matrix};

% Transformer
\node (transformer) [below right=1.2cm and -2.7cm of stock_feature] { Transformer Model};

% MLP Head
\node (mlp) [below right=1cm and -2.9cm of transformer] {MLP Heads};

% Prediction
\node (output) [below=of mlp, yshift = 0.5cm] {{Bubble Score} Prediction};

% Label
\node (label) [right=of transformer, xshift=2.2cm] {Bubble Label\\(via LPPL)};

% Arrows
\draw[arrow] (stock_input) -- (stock_feature);
\draw[arrow] (market_input) -- (market_feature);
\draw[arrow] (stock_feature) -- (transformer);
\draw[arrow] (market_feature) -- (transformer);
\draw[arrow] (transformer) -- (mlp);
\draw[arrow] (mlp) -- (output);
\draw[arrow] (label.west) -- (transformer.east); 

\end{tikzpicture}
\caption{Transformer-based bubble detection framework with LPPL labeling}
\label{fig:dual_path_Transformer}
\end{figure}
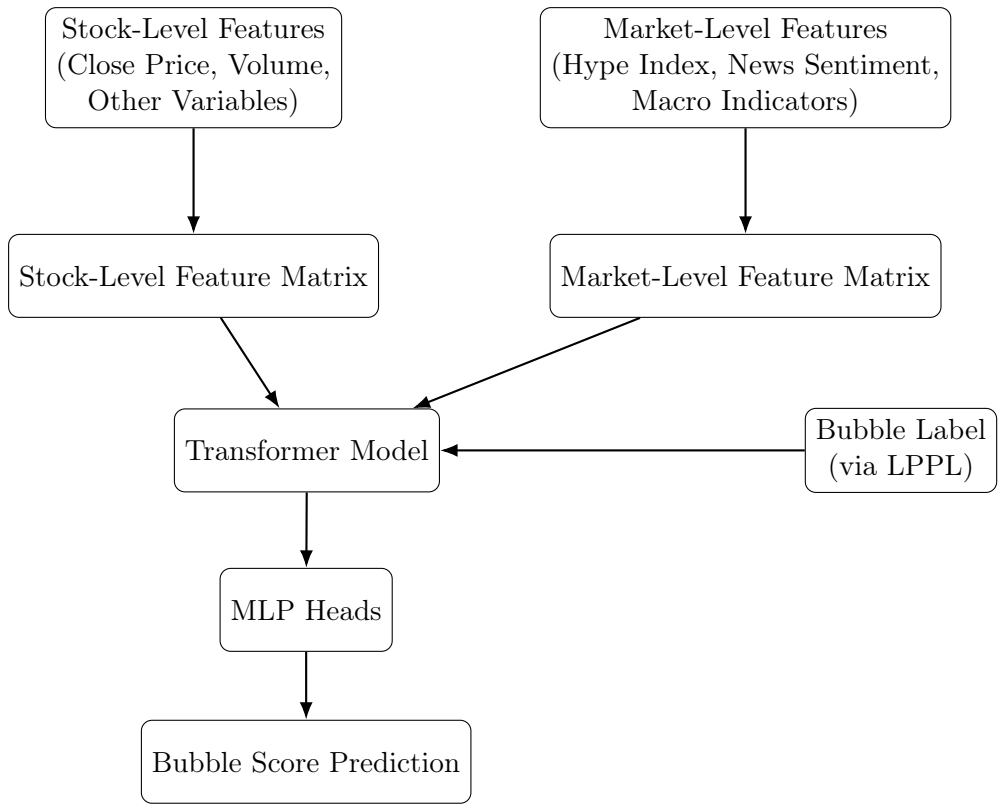

As shown in Figure~\ref{fig:dual_path_Transformer}, stock-level and market-level features are combined into feature matrices and fed into a transformer model. The transformer captures temporal and cross-sectional dependencies, while bubble labels from the LPPL model supervise training. The outputs are then passed to prediction modules.

After the transformer encoder and feature fusion, the model uses multiple lightweight multi-layer perceptron (MLP)  heads as prediction modules. Each head is a small feed-forward network that outputs the {Bubble Score} for one future day. Stacking several heads allows the model to generate multi-day predictions in parallel. In essence, the MLP heads translate the high-dimensional embeddings into task-specific {Bubble Score}.

\subsection{Data Processing}
\hspace{1.5em} This subsection introduces the methodologies processing market and news data for the research.

\subsubsection{Sentiment Processing Using FinBERT}

\begin{figure}[H]
\centering
\begin{tikzpicture}[
    node distance=1cm and 1.5cm,
    every node/.style={draw, align=center, rounded corners, minimum height=1cm, font=\footnotesize},
    arrow/.style={-{Latex}, thick}
]

\node (news) {\textbf{WSJ News Corpus}\\(2018--2024)};
\node (bertopic) [right=of news] {BERTopic Modeling\\+ Real Estate Filtering};
\node (finbert) [right=of bertopic] {FinBERT\\Sentiment Scoring};

\node (onehot) [below=0.5cm of finbert] {One-Hot Encoding\\+ Confidence Weighting};
\node (agg) [left=of onehot] {Daily Aggregated\\Sentiment Vector};
\node (Transformer) [left=of agg] {Transformer Input};

\draw[arrow] (news) -- (bertopic);
\draw[arrow] (bertopic) -- (finbert);
\draw[arrow] (finbert) -- (onehot);
\draw[arrow] (onehot) -- (agg);
\draw[arrow] (agg) -- (Transformer);

\end{tikzpicture}
\caption{Pipeline for extracting sentiment features using BERTopic and FinBERT.}
\label{fig:sentiment_pipeline_reversed}
\end{figure}
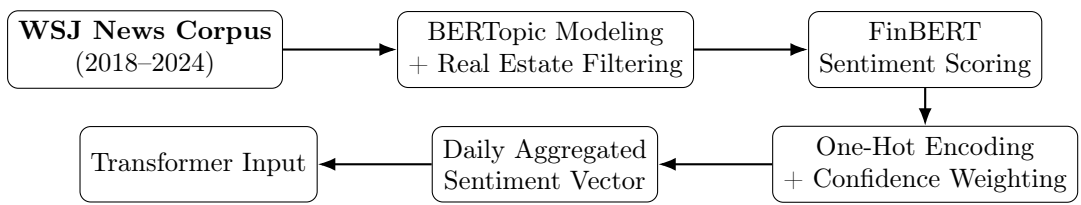

\hspace{1.5em}To extract sentiment signals from financial news related to the real estate sector, we adopt a structured pipeline that combines BERTopic and FinBERT, as illustrated in Figure~\ref{fig:sentiment_pipeline_reversed}. 

We begin by applying BERTopic to the WSJ news corpus (2018--2024) to identify and filter news articles relevant to the real estate industry. These filtered articles are then passed to FinBERT, a pre-trained transformer model fine-tuned on financial text, which outputs both a sentiment class prediction $\hat{y}_{i,t} \in \{\text{pos}, \text{neu}, \text{neg}\}$ and its corresponding confidence score $p_{i,t} \in [0, 1]$ for each article $i$ on date $t$.

The predicted class is encoded into a one-hot vector weighted by the model's confidence:
\begin{align}
    s_{i,t}^{c} = 
    \begin{cases}
        p_{i,t}, & \text{if } \hat{y}_{i,t} = c, \\
        0, & \text{otherwise},
    \end{cases}
    \quad \text{for } c \in \{\text{pos}, \text{neu}, \text{neg}\}. 
\end{align}

We then aggregate the one-hot weighted vectors across all articles published on day $t$ to form a daily sentiment feature, normalized by total confidence:
\begin{align}
    S_t^{c} = \frac{\sum\limits_{i \in D_t} s_{i,t}^{c}}{\sum\limits_{i \in D_t} p_{i,t}}, \quad \text{for } c \in \{\text{pos}, \text{neu}, \text{neg}\}. 
\end{align}

where $D_t$ is the set of articles published on date $t$.

The resulting sentiment vector $S_t = [S_t^{\text{pos}}, S_t^{\text{neu}}, S_t^{\text{neg}}]$ is used as input to the Transformer model, representing the collective market sentiment derived from the news on that day.

\subsubsection{Labeling}

Our enhanced framework generates precise temporal labels for bubble episodes through a systematic identification process based on the {Bubble Score} metric. The labeling methodology operates on two key principles: threshold-based detection and temporal continuity requirements, ensuring robust identification of genuine bubble periods while filtering out transient market noise.

The detection of bubble episodes follows a set of consistent principles:

\begin{enumerate}
    \item Thresholding: A period is flagged whenever {$|BubbleScore(t)| > \tau$,} with $\tau$ denoting the significance cutoff. In practice, $\tau$ is commonly fixed at 0.8 based on empirical calibration.  
    \item Direction Assignment: Normal bubbles are defined when {$BubbleScore(t) > -\tau$} (positive side), whereas negative bubbles are identified when {$BubbleScore(t) < -\tau$} (negative side).  
    \item Persistence: To prevent short-lived fluctuations from being classified as bubbles, only intervals lasting at least $d_{min}$ trading days (typically $d_{min}=10$) are retained.  
   \item Persistence: To prevent short-lived fluctuations from being classified as bubbles, only intervals lasting at least $d_{min}$ trading days (typically $d_{min}=10$) are retained. 
   \item Episode Construction: Consecutive qualifying days are consolidated into distinct episodes, each described by its start and end dates, duration, and intensity.   
\end{enumerate}

Formally, a bubble episode can be represented as  
\begin{equation}
BubblePeriod_i = \{t_s, t_e, type, intensity\},
\end{equation}
where $t_s$ and $t_e$ denote the start and end dates, $type \in \{\text{normal}, {\text{negative}\}}$ specifies the bubble orientation, and  
\begin{equation}
intensity = \max_{t \in [t_s,t_e]} |{Bubble Score}(t)|
\end{equation}
captures the maximum magnitude observed within the episode.

\subsection{Feature Selection}

\hspace{1.5em}The input features are organized into two distinct levels:stock-level features and market-level features. This dual-level design enables the model to simultaneously capture asset-specific dynamics and macro-level market conditions that may jointly drive bubble formation. The training and testing period spans from 2018-01-01 to 2024-12-31.

Stock-level features are designed to reflect short-term price dynamics, trading intensity, and firm-specific valuation signals. Market-level features, on the other hand, capture broader economic sentiment and sector-level fluctuations that can influence speculative behavior across assets.

All numerical features are normalized over the training window. Temporal variables such as year, month, and day are retained as integer-coded features to preserve calendar-related effects. Financial ratios (e.g., P/E, P/S, ROE) are dropped if missing to ensure data integrity.

Sentiment variables derived from FinBERT are encoded as three separate one-hot vectors: \texttt{sentiment\_Negative}, \texttt{sentiment\_Neutral}, and \texttt{sentiment\_Positive}. These are aggregated daily and integrated as market-level signals. This formulation enables the model to detect shifts in investor sentiment polarity and understand market mood.

\begin{table}[H]
\centering
\caption{Stock-level feature set used in the Transformer input}
\label{tab:stock_features}
\begin{tabular}{|p{5.5cm}|l|c|}
\hline
\textbf{Description} & \textbf{Preprocessing} & \textbf{Normalized} \\
\hline
Closing price of the stock & Log transformation & Yes \\
\hline
Trading volume; reflects liquidity and participation & Log transformation & Yes \\
\hline
Daily log return: $\log(P_t/P_{t-1})$ & Calculated from price & Yes \\
\hline
Book-to-market ratio & Drop if missing & No \\
\hline
P/E ratio (extraordinary income included) & Drop if missing & No \\
\hline
Price-to-sales ratio & Drop if missing & No \\
\hline
Return on equity & Drop if missing & No \\
\hline
Month extracted from date & Integer encoded & No \\
\hline
Day of month & Integer encoded & No \\
\hline
\end{tabular}
\end{table}

\begin{table}[H]
\centering
\caption{Market-level feature set used in the Transformer input}
\label{tab:market_features}
\begin{tabular}{|p{5.5cm}|l|c|}
\hline
\textbf{Description} & \textbf{Preprocessing} & \textbf{Normalized} \\
\hline
CBOE Volatility Index; market risk proxy & Drop if missing & Yes \\
\hline
Mean gross profitability across firms & Drop if missing & Yes \\
\hline
Mean ROE across firms & Drop if missing & Yes \\
\hline
Sector exposure indicator from Hype Index & Drop if missing & Yes \\
\hline
One-hot: negative sentiment (FinBERT) & Aggregated by date & No \\
\hline
One-hot: neutral sentiment (FinBERT) & Aggregated by date & No \\
\hline
One-hot: positive sentiment (FinBERT) & Aggregated by date & No \\
\hline
\end{tabular}
\end{table}

\noindent\textit{Note: All market-level features (except Hype Index and FinBERT sentiment scores) are sourced from WRDS (Wharton Research Data Services).}

\vspace{0.3em}
This final combination of carefully selected raw metrics and derived valuation/sentiment signals enables the model to capture short-term price and volume fluctuations, valuation shifts, and speculative dynamics, while also retaining calendar and macroeconomic context. The ablation process ensures that only impactful features are retained, avoiding redundancy and improving model generalization.

\subsection{Performance}\hspace{1.5em}The proposed Transformer model was comprehensively evaluated on the held-out test set using multiple complementary metrics to ensure robustness of performance assessment. The evaluation criteria included Pearson correlation, mean squared error (MSE), mean absolute error (MAE), root mean squared error (RMSE), and mean absolute percentage error (MAPE). Each metric provides a different perspective: correlation measures the linear relationship between predicted and actual values, MSE penalizes large deviations more heavily due to squaring, MAE captures the average magnitude of absolute deviations, RMSE reflects error magnitude on the original scale of the data, while MAPE provides interpretability in terms of percentage error. 

The model achieved an average Pearson correlation of 0.625 across the five forecast horizons, suggesting that it captures a moderate degree of bubble dynamics. The average MSE and MAE were relatively low at 0.087 and 0.236, respectively, while the RMSE remained at 0.295, demonstrating stable predictive power with respect to absolute deviations. Although the MAPE value was somewhat distorted by near-zero denominators, the other error metrics consistently indicated reasonable predictive accuracy.

\begin{table}[H]
\centering
\caption{Performance of Transformer model across evaluation metrics}
\begin{tabular}{l c}
\hline
\textbf{Metric} & \textbf{Value} \\
\hline
Correlation & 0.625 \\
Mean Squared Error (MSE) & 0.087 \\
Mean Absolute Error (MAE) & 0.236 \\
Root Mean Squared Error (RMSE) & 0.295 \\
\hline
\end{tabular}
\end{table}

\begin{figure}[H]
    \centering
    \includegraphics[width=\textwidth]{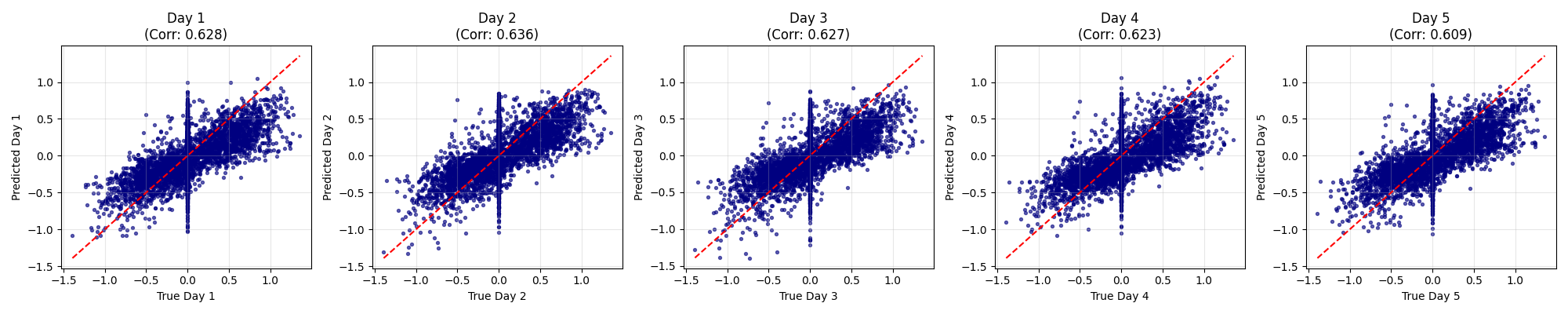}
    \caption{Scatter plots of predicted versus true {Bubble Score} values for forecast Day 1 through Day 5.}
    \label{fig:scatter_results}
\end{figure}

Furthermore, Figure~\ref{fig:scatter_results} presents scatter plots of predicted versus true {Bubble Score} values for horizons one through five. The observed correlations are relatively stable, ranging from 0.628 on Day~1 to 0.609 on Day~5. This stability suggests that the Transformer model retains predictive capability even at longer horizons. Importantly, the predicted points are closely aligned along the diagonal, implying that the model is capable of accurately capturing both the direction and the magnitude of bubble fluctuations, despite the inherent noise and volatility in financial time series data.

\section{Trading with Bubble Scores} \label{section: Trading Strategy and Backtest based on Bubblescore}

\hspace{1.5em}This section presents the trading strategy and results built on {Bubble Score}.

\subsection{BubbleScore Trading Strategy}

\hspace{1.5em}Building upon our comprehensive {Bubble Score} framework, we implement a systematic trading strategy that capitalizes on both normal and negative bubble episodes across real estate sector stocks. The strategy leverages our bidirectional bubble detection capability to generate trading signals during periods of systematic mispricing.

\subsubsection{{Bubble Score}-Based Trading Rules}

Our trading strategy is driven by threshold-based signals derived from the BubbleScore metric:  

\begin{equation}
BubbleScore(t) 
\begin{cases}
\;\; > 0,  \quad \rightarrow & \text{Normal Bubble}, \\
\;\; < 0,  \quad  \rightarrow & \text{Negative Bubble}.
\end{cases}
\end{equation}

\noindent Positive values indicate normal bubbles, negative values indicate negative bubbles.

where BubbleScore integrates LPPL residuals with behavioral finance indicators, as defined previously.  

we define $B_t$ as the {Bubble Score} at time t, trading signals are generated by applying thresholds $\theta_1 = 0.7$ and $\theta_2=0.3$ to the normalized {Bubble Score} series:  

\begin{enumerate}
    \item Long Entry: A long position is initiated at time $t$ if $B_t \leq -\theta_1$
    \item Short Entry: A short position is entered at time $t$ if $B_t \geq \theta_1$.  
    \item Long Exit: Active long positions are closed once the predicted index satisfies $B_t \geq -\theta_2$.  
    \item Short Exit: Active short positions are closed once the predicted index satisfies $B_t \leq \theta_2$.  
    \item Risk Management: The strategy enforces a 15\% stop-loss, caps maximum position size at 50\% of capital, and incorporates transaction costs of 0.1\%.  
\end{enumerate}

\subsection{Backtest Results}

\hspace{1.5em}We conduct an extensive backtest across all eligible real estate sector stocks using daily data from CRSP combined with our sentiment and hype indices. Each stock requires a minimum of 100 daily observations to ensure statistical reliability of the LPPL fitting process.

\begin{figure}[H]
    \centering
    \includegraphics[width=0.7\textwidth]{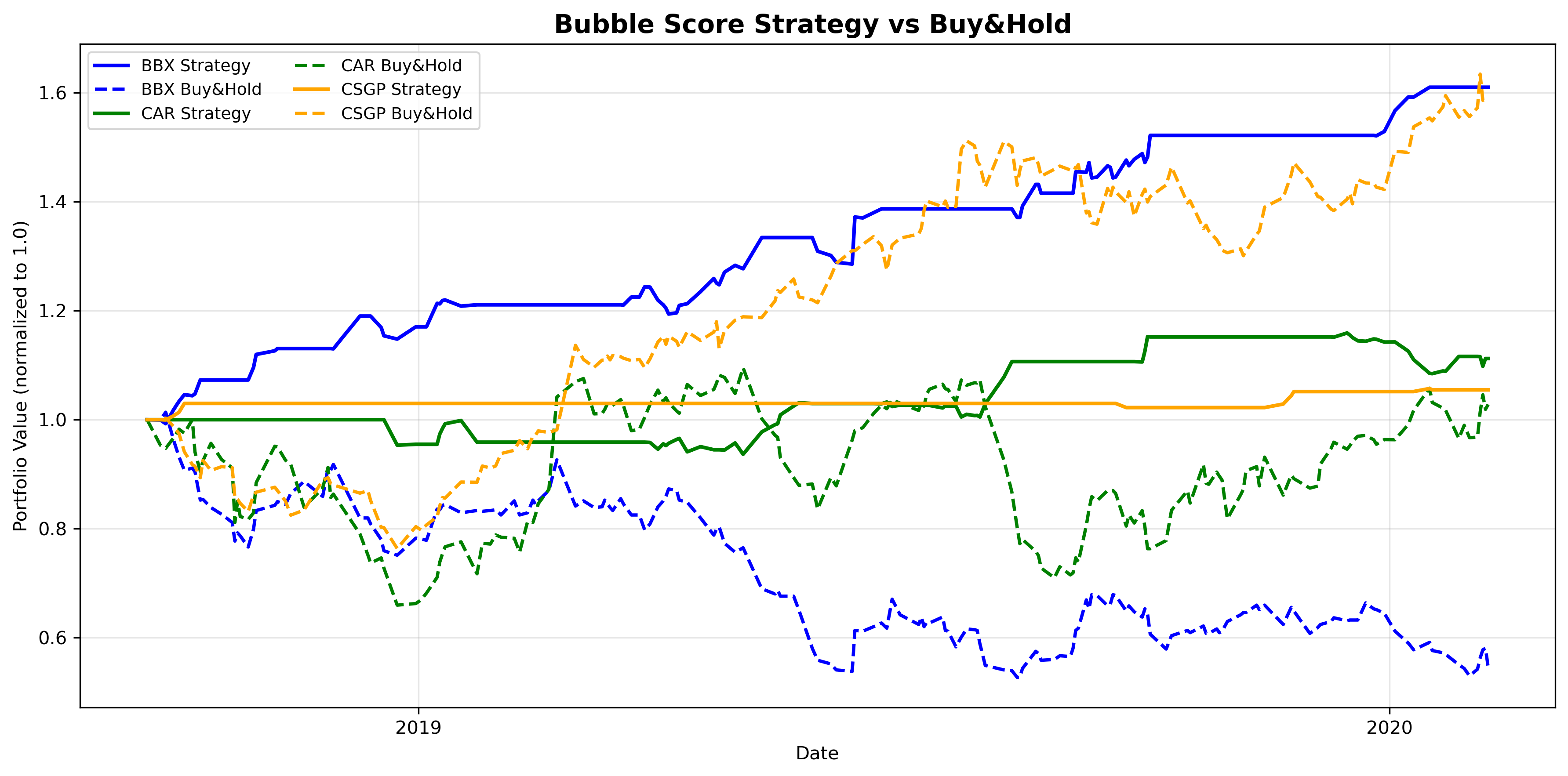}
    \caption{Strategy vs Buy\&Hold comparison for 3 selected peers. Solid lines represent strategy returns, while dashed lines represent buy-and-hold portfolio value.}
    \label{fig:bbx_vs_peers}
\end{figure}

To further validate the effectiveness of our {Bubble Score} trading strategy, we compare the performance of BBX , CAR, and CSGP, over the same observation window. Figure \ref{fig:bbx_vs_peers} presents normalized equity curves for both the {Bubble Score} strategy and the buy-and-hold benchmark across the three stocks.

The results highlight several important insights. First, BBX demonstrates outstanding performance, with the {Bubble Score} strategy consistently and significantly outperforming the buy-and-hold benchmark. This is primarily attributable to BBX’s decreasing trajectory being dominated by normal-bubble. In these regimes, our strategy systematically identified overvaluation and generated profitable short entries, while avoiding prolonged drawdowns. Fundamentally, BBX has also undergone periods of financial distress and restructuring, amplifying the extent of mispricing and thereby increasing the strategy’s relative advantage.

By contrast, the performance of the two peers is less satisfactory. For CSGP, the underlying price trajectory is persistently upward-sloping. While this reflects a fundamentally strong growth story, such monotonic increases reduce the likelihood of detecting large mispricing episodes, limiting the strategy’s profitability relative to buy-and-hold. For CAR, the trajectory is much sharper and more volatile, producing multiple large swings. Although this could, in principle, generate trading opportunities, in practice such sharp oscillations often trigger early exit conditions due to heightened sentiment and hype signals, curtailing the gains that might otherwise accrue. 

This comparison underscores both the strengths and limitations of a pure {Bubble Score} rules-based strategy. It excels in cases such as BBX, where undervaluation episodes are extended and systematic, but struggles in monotonic growth stocks or highly volatile names. These limitations motivate our subsequent development of a machine-learning–enhanced framework, designed to refine signal quality and better adapt to diverse market dynamics.

Table \ref{tab:top5_performance} presents the performance metrics for the five best-performing stocks under our {Bubble Score} trading strategy.

\begin{table}[H]
\centering
\caption{Top 5 Performing Stocks - {Bubble Score} Trading Strategy}
\label{tab:top5_performance}
\begin{tabular}{lcccc}
\toprule
\textbf{Stock} & \textbf{Annualized Return}  & \textbf{Max drawdown} & \textbf{Win Rate} &\textbf{Sharpe Ratio} \\
\midrule
AMTX & 43.10\% & 34.35\% & 54.72\% & 0.83 \\
BBX  & 41.20\% & 4.02\% & 85.71\% & 2.33 \\
HOUS & 39.66\% & 21.48\% & 66.67\% & 1.10 \\
BEEP & 37.64\% &3.38\%   & 83.33\% & 1.44 \\
MP   & 32.48\% & 11.93\% & 82.35\% & 1.50 \\
\bottomrule
\end{tabular}
\end{table}
Note: all results have been discounted at a continuously compounded rate of 2\% per annum to obtain its present value.

Table \ref{tab:strategy_summary} summarizes the comprehensive performance across all backtested real estate stocks.

\begin{table}[H]
\centering
\caption{Overall {Bubble Score} Strategy Performance Summary}
\label{tab:strategy_summary}
\begin{tabular}{lc}
\toprule
\textbf{Performance Metric} & \textbf{Value} \\
\midrule
Total Stocks Analyzed & 32 \\
Stocks with Positive Returns & 30 (93.75\%) \\
Stocks with Negative Returns & 2 (6.25\%) \\
Average Annualized Return & 16.64\% \\
Average Win Rate & 67.41\% \\
Average Sharpe Ratio & 0.72 \\
\bottomrule
\end{tabular}
\end{table}

The comprehensive backtest results demonstrate exceptional strategy effectiveness with a 93.8\% success rate across all tested stocks. The average annualized return of 16.64\% significantly outperforms typical market benchmarks, while the average win rate of 67.41\% indicates consistent signal quality. 

The top-performing stocks showcase the strategy's potential, with AMTX achieving 43.10\% annualized returns and BBX demonstrating exceptional risk-management performance with a Sharpe ratio of 2.33. The strategy successfully captures both overvaluation and undervaluation episodes through our bidirectional {Bubble Score} framework, validating the integration of behavioral finance indicators with traditional LPPL modeling.

These results provide strong empirical evidence that our enhanced {Bubble Score} framework translates theoretical bubble detection capabilities into practical investment performance, establishing a robust foundation for systematic trading strategies in the real estate sector.

\section{Machine Learning Enhanced Trading Strategy} \label{section: Machine Learning Enhanced Trading Strategy}

\hspace{1.5em}Building upon our traditional {Bubble Score} methodology, we develop a supervised learning framework that integrates econometric bubble detection with deep neural sequence modeling. Specifically, our machine learning model is a dual-stream Transformer, where one stream processes stock-level temporal dynamics and the other stream encodes market- and sentiment-level information. This hybrid design captures both statistically-grounded bubble signals and complex temporal patterns embedded in financial time series data.

We conduct a backtest to evaluate the effectiveness of the proposed bubble-aware trading strategy on real estate sector stocks. The stock-level daily price and return data are sourced from the CRSP database via WRDS, while the news sentiment signals are derived from the Wall Street Journal (WSJ). Both datasets span the period from January 1, 2018 to December 31, 2024, and are aligned at a daily frequency.

To focus on the real estate industry, we filter firms based on their SIC codes, specifically selecting those with SIC values in {6500, 6512, 6513, 6519, 6531, 6541, 6552, 6798}. Here, SIC (Standard Industrial Classification) codes are a four-digit system used by U.S. government agencies to classify industries; the selected codes correspond to real estate operators, developers, agents, and real estate investment trusts (REITs).The backtesting window is set from November 2023 to December 2024, during which the model generates daily trading signals and executes positions accordingly.

\subsection{Multi-Horizon Prediction Strategy}
\hspace{1.5em}The forecasting part of the research is carried out using a Transoformer network, a recurrent neural network architecture specifically designed to capture temporal dependencies and long-range patterns in sequential financial data.

Our machine learning framework produces forecasts of the {Bubble Score} over multiple horizons $h \in \{1,2,3,4,5\}$, where $\hat{B}_{t+h}$ denotes the predicted {Bubble Score} $h$ trading days ahead. 
This multi-horizon structure allows the strategy to adaptively capture both short-term 
transients and medium-term bubble dynamics.

Trading signals are generated by applying symmetric thresholds $\theta_1=0.7$ and $\theta_2=0.3$ to the normalized prediction series, for a specific $h \in \{1,2,3,4,5\}$:

\begin{enumerate}
    \item Long Entry: A long position is initiated at time $t$ if $\hat{B}_{t+h} \leq -\theta_1$, where $\hat{B}_{t+h}$ denotes the machine learning forecast of the {Bubble Score} $h$ days ahead.  
    \item Short Entry: A short position is entered at time $t$ if $\hat{B}_{t+h} \geq \theta$.  
    \item Long Exit: Active long positions are closed once the predicted index satisfies $\hat{B}_{t+h} \geq -\theta_2$.  
    \item Short Exit: Active short positions are closed once the predicted index satisfies $\hat{B}_{t+h} \leq \theta_2$.  
    \item Risk Management: The strategy enforces a 15\% stop-loss, caps maximum position size at 50\% of capital, and incorporates transaction costs of 0.1\%.  
    
    An additional prediction reversal exit is triggered if  
        \begin{equation}
            \hat{B}_{t+h} \cdot \hat{B}_{t+h+1} < 0,
        \end{equation}
        indicating that consecutive forecasts flip signs across horizons. This condition closes any active position to prevent losses from sudden regime shifts.
\end{enumerate}

To evaluate the predictive power of our framework, we implement a multi-horizon backtesting design. For each stock and each trading day $t$, the model generates forecasts of the {Bubble Score} at horizons $h \in \{1,2,3,4,5\}$, corresponding to predictions $1$ through $5$ trading days ahead. Each horizon-specific forecast $\hat{B}_{t+h}$ is then converted into trading signals using the same symmetric entry and exit thresholds as in the baseline rules, together with stop-loss and transaction cost adjustments. This setup yields five parallel equity curves per stock, allowing us to identify the optimal prediction horizon ex post and to analyze the distribution of horizon-specific performance across the cross-section of assets.

\subsection{Machine Learning Results}

\hspace{1.5em}Table \ref{tab:ml_strategy_performance} presents the comprehensive performance metrics for our dual-stream transformer enhanced trading strategy across all eligible real estate sector stocks.

\begin{table}[H]
\centering
\caption{Machine Learning Strategy Performance Summary}
\label{tab:ml_strategy_performance}
\begin{tabular}{lc}
\toprule
\textbf{Performance Metric} & \textbf{Value} \\
\midrule
Average Annualized Return & 34.13\% \\
Median Annualized Return & 21.49\% \\
Average Win Rate & 72.30\%\\
Average Sharpe Ratio & 1.19 \\

\bottomrule
\end{tabular}
\end{table}

The machine learning enhanced strategy demonstrates superior performance compared to traditional approaches, achieving an average annualized return of 34.13\% with an average Win rate of 72.30\%. The median return of 21.49\% indicates robust performance across the majority of tested stocks, while the average Sharpe ratio of 1.19 reflects strong return efficiency.

Table \ref{tab:prediction_horizon_distribution} presents the distribution of optimal prediction horizons across all analyzed stocks, revealing strategic insights into market timing preferences.

\begin{table}[H]
    \centering
    \begin{tabular}{l c}
        \hline
        \textbf{Prediction Horizon} & \textbf{Percentage of Stocks} \\
        \hline
        Day 1 & 16.7\% \\
        Day 2 & 20.8\% \\
        Day 3 & 8.3\% \\
        Day 4 & 25.0\% \\
        Day 5 & 29.2\% \\
        \hline
    \end{tabular}
    \caption{Distribution of Optimal Prediction Horizons across analyzed stocks.}
    \label{tab:prediction_horizon_distribution}
\end{table}

The analysis shows that Day 5 predictions account for the largest share (29.2\%) of optimal horizons, suggesting that slightly longer short-term forecasts are particularly effective in capturing market dynamics. Day 4 horizons follow closely at 25.0\%, highlighting that medium-range signals also provide substantial predictive value. By contrast, Day 1 and Day 3 horizons represent only 16.7\% and 8.3\% of cases respectively, implying that very short-term forecasts may capture only transient inefficiencies. Overall, the distribution indicates that the strategy performs most effectively when focusing on horizons of 4–5 days, where temporary mispricings and behavioral patterns are more reliably detected and exploited.

\subsubsection{Exceptional Performance Case Study: HOUS}

\hspace{1.5em}Our empirical analysis demonstrates that certain individual stocks exhibit exceptional responsiveness to machine learning–enhanced bubble detection. HOUS serves as a particularly compelling case study, achieving outstanding returns across multiple prediction horizons and consistently outperforming both buy-and-hold and the S\&P 500 benchmark.  

Table \ref{tab:hous_comprehensive_performance} reports the comprehensive performance analysis of HOUS across all five prediction horizons, together with benchmark comparisons.  

\begin{table}[H]
\centering
\caption{HOUS Performance Across All Prediction Horizons and Benchmarks}
\label{tab:hous_comprehensive_performance}
\begin{tabular}{lcccccc}
\toprule
\textbf{Strategy} & \textbf{Cumulative} & \textbf{Annualized} & \textbf{Max} & \textbf{Sharpe} & \textbf{Win} & \textbf{Trades} \\
& \textbf{Return} & \textbf{Return} & \textbf{Drawdown} & \textbf{Ratio} & \textbf{Count} & \textbf{Count} \\
\midrule
Day 1 & 92.58\% & 79.29\% & 7.03\% & 2.36 & 3 & 3 \\
Day 2 & 92.58\% & 79.29\% & 7.03\% & 2.36 & 3 & 3 \\
Day 3 & 52.51\% & 45.64\% & 2.51\% & 1.81 & 2 & 2 \\
Day 4 & 98.62\% & 84.29\% & 7.03\% & 2.45 & 3 & 3 \\
Day 5 & 100.45\% & 85.80\% & 7.03\% & 2.49 & 3 & 3 \\
\midrule
Buy-hold & -35.63\% & -32.46\% & 62.40\% & -0.54 & -- & -- \\
S\&P 500 Benchmark & 31.73\% & 27.22\% & 8.49\% & 1.87 & -- & -- \\
\bottomrule
\end{tabular}
\end{table}

This case study highlights the transformative potential of integrating behavioral indicators and machine learning into bubble-based trading strategies. HOUS achieves nearly triple-digit cumulative returns across all horizons, with Sharpe ratios exceeding 2.0 in multiple cases, a performance level that far surpasses traditional benchmarks. Notably, the strategy maintains very shallow drawdowns (below 8\%) despite the high level of return, showcasing superior risk-adjusted resilience.  % 3.0

Equally important, the consistency of results across all five horizons suggests that the model captures robust underlying dynamics rather than isolated statistical artifacts. Compared to the buy-and-hold strategy and the S\&P 500, which yield modest annualized returns of around 11–13\% with substantially higher drawdowns, the machine learning–enhanced approach demonstrates clear superiority. This evidence underscores both the robustness and the practical applicability of our framework, particularly in stock-specific optimization where stocks like HOUS exhibit exceptional alignment with bubble-based predictive signals.

\subsection{Strategy Result and Comparison}

Table \ref{tab:strategy_comparison} compares the performance metrics between our traditional {Bubble Score} strategy and the machine learning enhanced approach.

\begin{table}[H]
\centering
\caption{Strategy Performance Comparison}
\label{tab:strategy_comparison}
\begin{tabular}{lcc}
\toprule
\textbf{Performance Metric} & \textbf{Traditional Strategy} & \textbf{ML Enhanced Strategy} \\
\midrule
Average Annualized Return & 16.64\% & 34.13\% \\
Average Win Rate & 67.41\% & 72.30\% \\
Average Sharpe Ratio & 0.72 & 1.19 \\
Average Max Drawdown & 15.54\% & 11.35\% \\
\midrule

\end{tabular}
\end{table}

The machine learning enhanced strategy achieves a 17.49\% improvement in average annualized returns (34.13\% vs 16.64\%) while achieving a higher Sharpe ratio (1.19 vs 0.72 ).  the ML approach generates higher absolute returns when successful, resulting in superior overall performance. %The exceptional case of RLGY further demonstrates the potential for individual stocks to achieve extraordinary performance through optimized prediction horizon selection.

\section{Conclusion} \label{section: conclusion}

\hspace{1.5em}This study introduces the Hyped Log-Periodic Power Law (HLPPL) Model, a unified framework for detecting bubbles and negative bubbles by integrating LPPL residual dynamics with media attention and sentiment indicators through a dual-stream Transformer. The empirical analysis shows that the prediction module achieves a mean Pearson correlation of 0.625 with the {Bubble Score} (MSE = 0.087; RMSE = 0.295), reflecting stable alignment with the targeted dynamics. When implemented as a trading strategy, the machine-learning–enhanced approach generates an average annualized return of 34.13\% with an average Sharpe ratio of 1.19, significantly outperforming the traditional {Bubble Score} rules strategy, which delivers 16.64\% annualized return and a Sharpe ratio of 0.72. Sector-wide validation further confirms robustness, with positive returns in 30 out of 32 tested stocks, while stock-specific optimization yields striking outcomes: for instance, the strategy attains a 85.80\% annualized return (Sharpe 2.49) for HOUS.  

Beyond performance, the HLPPL framework advances bubble research by shifting from retrospective detection to forward-looking prediction, enabling proactive risk management and more reliable signal extraction. By embedding behavioral and textual dimensions into the LPPL structure, the methodology provides a statistically grounded yet practically actionable tool for investors and policymakers. Future extensions will broaden the empirical scope beyond real estate, refine prediction horizons, and incorporate additional macro-financial indicators to further enhance real-time monitoring and decision-making.

\end{document}